\documentstyle[aps,prb,amsmath,amssymb,epsfig]{revtex}
\begin{document}
\title{Parquet approach to nonlocal vertex functions and electrical
  conductivity of disordered electrons}

\author{V.  Jani\v{s}}

\address{Institute of Physics, Academy of Sciences of the Czech Republic,\\
  Na Slovance 2, CZ-18221 Praha 8, Czech Republic\\ e-mail:janis@fzu.cz}

\date{\today}

\maketitle

\begin{abstract}
  A diagrammatic technique for two-particle vertex functions is used
  to describe systematically the influence of spatial quantum
  coherence and backscattering effects on transport properties of
  noninteracting electrons in a random potential. In analogy with
  many-body theory we construct parquet equations for topologically
  distinct {\em nonlocal} irreducible vertex functions into which the
  {\em local} one-particle propagator and two-particle vertex of the
  coherent-potential approximation (CPA) enter as input. To complete
  the two-particle parquet equations we use an integral form of the
  Ward identity and determine the one-particle self-energy from the
  known irreducible vertex.  In this way a conserving approximation
  with (Herglotz) analytic averaged Green functions is obtained. We
  use the limit of high spatial dimensions to demonstrate how nonlocal
  corrections to the $d=\infty$ (CPA) solution emerge. The general
  parquet construction is applied to the calculation of vertex
  corrections to the electrical conductivity. With the aid of the
  high-dimensional asymptotics of the nonlocal irreducible vertex in
  the electron-hole scattering channel we derive a mean-field
  approximation for the conductivity with vertex corrections.  The
  impact of vertex corrections onto the electronic transport is
  assessed quantitatively within the proposed mean-field description
  on a binary alloy.
\end{abstract}

%

\section{Introduction}
\label{sec:intro}

Randomness in the chemical composition or due to defects in solids
causes elastic scatterings of charge carriers that influence
significantly their motion and reduce their mobility. To describe the
effects of randomly distributed scatterers reliably one has to use
approximations taking satisfactorily into account quantum coherence
between scattering events. In particular, self-consistence between one
and two-particle functions is needed if we want to assess the role of
backscatterings and the backflow on transport properties of electrons
in a random potential.

A simplest possibility to account for quantum coherence is to sum all
multiple scatterings on a single scatterer. A self-consistent theory with
all single-site scatterings is the Coherent-Potential Approximation (CPA)
and was developed in the end of sixties.\cite{Soven67,Velicky68}  It is a
mean-field approximation for disordered (noninteracting) electrons. It
provides a local coherent potential (self-energy) that, in the
thermodynamic limit, comprises the effects of the random potential on the
motion of the single electron. The CPA, as other mean-field theories,
however, suppresses spatial coherence between distinct scattering centers
and the moving electrons feel the influence of the random potential only
via an averaged medium.

CPA has proven very successful in the description of one-electron
properties at the model level as well as in realistic calculations of
random alloys.\cite{Elliot74,Turek97} Although this approximation can as
well describe two-particle averaged functions, due to the lack of spatial
coherence, it fails to capture backscattering effects on the transport
coefficients and the electrical conductivity. The two-particle CPA vertex
does not depend on the transfer momentum between the incoming and the
outgoing particles and hence the CPA conductivity reduces in the
single-band bulk systems to a contribution from a single particle-hole
bubble.  There are no vertex corrections to the electrical conductivity
within the standard CPA.\cite{Velicky69}

Vertex corrections to the single-bubble one-electron conductivity are
important in various situations. In low dimensions ($d\le2$) or for
sufficiently strong disorder they lead to Anderson
localization.\cite{Lee85} Further on, tunnel conductance or transport
through multilayered dirty metals are essentially influenced by vertex
corrections.\cite{Itoh99} To obtain more realistic results for the
electronic transport in dirty metals one has to go beyond the standard CPA
to the conductivity and develop approximations containing spatial quantum
coherence and backscattering contributions.

There is a long history of efforts to improve upon the mean-field CPA
description of disordered electrons \cite{Gonis92}. Most of them
concentrate on single-particle properties and improve upon the CPA in the
self-energy (coherent potential). A natural extension of the single-site
theory is to use clusters self-consistently embedded in an averaged medium.
However, apart from the traveling-cluster
approximation,\cite{Mills78,Kaplan80} extensions in the lattice space fail
to warrant global analytic properties of the solution and hence spurious
effects can emerge.\cite{Capek72,Nickel73} Only recently a cluster
expansion in momentum space was suggested that warrants analytic (Herglotz)
properties of the resulting averaged propagators and the self-energy at
each stage.\cite{Jarrell00} Cluster approximations with self-energy
diagrams improve also two-particle vertex functions.  However, cluster
approximations reduce spatial quantum coherence only to a discrete set of
lattice sites or momenta. Such approximations then remain perturbative in
the coherence range and cannot lead to Anderson localization to which we
need long-range coherence with infinite-many backscattering or ``crossed''
diagrams.\cite{Vollhardt92,Kramer93}

Using cluster approximations to improve upon the mean-field transport
properties means that we have first to extend the one-electron calculation
scheme. A tremendous effort at the one-particle level is to be exerted to
obtain significant changes in transport properties. Cluster expansions are
hence not very effective in calculating quantum coherence effects in the
electrical conductivity. It is more efficient to develop approximations
directly for two-particle functions.

A suitable framework for developing two-particle approximations is a
parquet approach devised within quantum many-body
theory.\cite{Dominicis62,Jackson82,Bickers91,Janis99b} It is an advanced
scheme of summation of Feynman (many-body) diagrams based on
renormalizations of two-particle vertex functions. Its main idea is to
utilize ambiguity in the definition of the two-particle irreducibility.
Each two-particle irreducibility, i.~e., the way how pairs of one-particle
propagators are cut without disconnecting the diagram, defines a scattering
channel and a Bethe-Salpeter equation for the full two-particle vertex.
Since different two-particle channels are topologically inequivalent, a
solution of the Bethe-Salpeter equation from one channel (reducible
function) is irreducible in the other channels where it is used in the
input (integral kernel) in the respective Bethe-Salpeter equations. Thereby
a set of coupled, nonlinear self-consistent equations for the two-particle
irreducible vertices (parquet equations) is obtained.  Parquet equations
have been applied onto various many-body problems, but no significant
attempt has been made to use the parquet-type renormalization of Feynman
diagrams in disordered systems.

In this paper we develop a parquet approach to systems with noninteracting
electrons subjected to a random potential. We show how to construct
controlled approximations directly for the two-particle vertex using the
idea of parquet diagrams. Since the parquet construction applies only to
nonlocal propagators, we start from the limit of high spatial dimensions
where the diagonal and off-diagonal one-particle propagators separate and
the CPA becomes exact.\cite{Note1} Beyond this limit we construct parquet
equations for two-particle irreducible vertices from Bethe-Salpeter
equations with a perturbed nonlocal one-particle propagator and the local
two-particle vertex as input. Next we use a Ward identity to determine the
self-energy and the full one-particle averaged propagator from the
calculated vertex functions. This self-consistent procedure warrants
conservation laws and analytic properties of the one-particle functions
whenever the solutions to the two-particle parquet equations are analytic.

The unrestricted system of parquet equations is not soluble in general. We
hence resort to high spatial dimensions where the two-particle
self-consistence is naturally suppressed and one obtains the asymptotic
form of the two-particle vertex in closed form.\cite{Janis99a} We use this
explicit result to derive a mean-field approximation for the electrical
conductivity containing vertex corrections. We then choose a binary alloy
to make quantitative assessments of the impact of vertex corrections on the
bulk conductivity.

The paper is organized as follows. We derive in Sec.~\ref{sec:vertex_hd}
the parquet equations for vertex functions of disordered electrons. In
Sec.~\ref{sec:Ward_identity} we show how Ward identities can be used to
determine the self-energy from a given irreducible vertex function so that
we preserve conserving character of the approximation. The electrical
conductivity with the irreducible vertex function in the electron-hole
scattering channel is derived in Sec.~\ref{sec:conductivity} where we use
the result to constructing a mean-field approximation for the electrical
conductivity with vertex corrections. The mean-field approximation is
applied in high spatial dimensions on a binary alloy in
Sec.~\ref{sec:binary_alloy} to obtain quantitative results.

\section{Calculation of the vertex function}
\label{sec:vertex_hd}

We use a tight-binding Anderson model of noninteracting spinless electrons
moving in a random, site-diagonal potential $V_i$ described by a
Hamiltonian
\begin{eqnarray}\label{eq:AD_hamiltonian}
\widehat{H}_{AD}&=&\sum_{<ij>}t_{ij}c_{i}^{\dagger} c_{j}+\sum_iV_ic_{i
  }^{\dagger } c_{i} \, .
\end{eqnarray}
The values of the random potential $V_i$ are site-independent and obey a
disorder distribution $\rho(V)$. I.~e., a function depending on the random
potential $V_i$ is averaged via
\begin{eqnarray}   \label{eq:averaging}   \left\langle
X(V_i)\right\rangle_{av}&=&\int_{-\infty}^{\infty}dV\rho(V)X(V)\, .
\end{eqnarray}

Solving the problem of disordered electrons in thermodynamic equilibrium
amounts to finding the averaged free energy defined as
\begin{eqnarray}\label{av_free_energy} F_{av}&=&-k_BT\left\langle \ln
\mbox{Tr}\exp \left\{ -\beta \widehat{H}_{AD}(t_{ij,}V_i)\right\}
\right\rangle _{av}\, ,
\end{eqnarray}
where the trace $\mbox{Tr}$ runs over the electronic degrees of freedom in
the Fock space. However, the averaged free energy does not contain the
entire information about the disordered system. In particular we cannot
derive transport properties and the response to disturbing external forces
from it. We need to know averaged products of Green functions for different
energies. To include external perturbations into the thermodynamic
description we introduce a new quantity
$\Omega^{\nu}(E_1,E_2,\hdots,E_\nu;U)$. It is a generalized averaged grand
potential with $\nu$ energy states coupled via an external perturbation
$U$. We define
\begin{equation}
  \label{eq:N_grand_potential}
  \Omega^{\nu}(E_1,E_2,\hdots,E_\nu;U)=-k_BT\left\langle\ln\mbox{Tr}\exp
  \left\{-\beta\sum_{i,j=1}^\nu\left(\widehat{H}^{(i)}_{AD}\delta_{ij}-E_i
      \widehat{N}^{(i)}\delta_{ij} + \Delta\widehat{H}^{(ij)}
  \right)\right\} \right\rangle_{av}
\end{equation}
where we assigned to each (complex) energy $E_i$ a separate Hilbert
state space and
$\Delta\widehat{H}^{(ij)}=\sum_{kl}U^{(ij)}_{kl}c_{k}^{(i)\dagger
  }c^{(j)}_{l}$. Potential $\Omega^{\nu}(E_1,E_2,\hdots,E_\nu;U)$ is a
generating functional for averaged products of Green functions up to the
$\nu$th order. In practice, within linear-response theory we will use only
one and two-particle Green functions, i.~e.,
$\Omega^{\nu}(E_1,E_2,\hdots,E_\nu;U)$ is expanded up to $U^2$.

Averaged Green functions (propagators) are fundamental quantities with the
aid of which we can calculate all characteristics of the disordered system.
We can use momenta as good quantum numbers, since translational invariance
is restored for the averaged quantities. Averaged propagators can then be
expressed as sums of Feynman diagrams for disordered systems in analogy to
the standard many-body diagrams.\cite{Note2}

The averaged one-particle propagator is represented with the aid of the the
self-energy or a coherent potential $\Sigma({\bf k},z)$ that comprises the
influence of fluctuations of the random potential onto the motion of the
single electron. We write
\begin{eqnarray}   \label{eq:av_1PP}
G({\bf k},z)&=&\frac 1{z-\epsilon({\bf k})-\Sigma({\bf k},z)} =\frac
1N\sum_{ij}e^{-i{\bf k}({\bf R}_i-{\bf R}_j)}
\left\langle\left[z\widehat{1}-\widehat{t}-\widehat{V}\right]^{-1}_{ij}
\right\rangle_{av}
\end{eqnarray}
where the first equality expresses the Dyson equation connecting the
irreducible one-particle function (self-energy) with the one-particle
averaged propagator.

The averaged two-particle propagator is defined as
\begin{eqnarray}
  \label{eq:av_2PP}
  G^{(2)}_{ij,kl}(z_1,z_2)&=&\left\langle\left[z_1\widehat{1}-\widehat{t}
      -\widehat{V}\right]^{-1}_{ij}  \left[z_2\widehat{1}-\widehat{t}
      -\widehat{V}\right]^{-1}_{kl} \right\rangle_{av}
\end{eqnarray}
to which we define the Fourier transform to momentum space as follows
\begin{eqnarray}
  \label{eq:2P_momentum}
   G^{(2)}({\bf k}_1,z_1,{\bf k}_2,z_2;{\bf q})&=& \frac
      1{N}\sum_{ijkl}e^{-i{\bf k}_1{\bf R}_i} e^{i({\bf k}_1+{\bf
      q}){\bf R}_j} e^{-i({\bf k}_2+{\bf q}){\bf
      R}_k} e^{i{\bf k}_2{\bf R}_l} G^{(2)}_{ij,kl}(z_1,z_2)\, .
\end{eqnarray}
The two-particle Green function $G^{(2)}({\bf k}_1,z_1,{\bf k}_2,z_2;{\bf
  q})$ contains also uncorrelated motion of two separate particles. The
actual measure of a correlated motion of two particles is a vertex function
\begin{eqnarray}
  \label{eq:2P_vertex}
  \Gamma({\bf k}_1,z_1,{\bf k}_2,z_2;{\bf q})&=& G^{-1}({\bf k}_1,z_1)
G^{-1}({\bf k}_2,z_2)\left[G^{(2)}({\bf k}_1,z_1,{\bf k}_2,z_2;{\bf q})
\right.\nonumber\\&&\left. -\delta({\bf q})G({\bf k}_1,z_1) G({\bf k}_2,z_2)
\right] G^{-1}({\bf k}_1+{\bf q},z_1) G^{-1}({\bf k}_2+{\bf q},z_2)\, .
\end{eqnarray}

In analogy to the Dyson equation for the one-particle propagator we can
represent the two-particle vertex with the aid of a two-particle
irreducible vertex $\Lambda$ and a Bethe-Salpeter equation. Unlike the
one-particle case, the Bethe-Salpeter equation is not defined unambiguously
whenever we work with nonlocal propagators. This fact we utilize later in
the parquet construction. For the present moment we take the Bethe-Salpeter
equation in the electron-hole channel describing multiple scatterings of a
pair of an electron and a hole and write
\begin{eqnarray}
  \label{eq:eh_BS}
  \Gamma({\bf k}_1,z_1,{\bf k}_2,z_2;{\bf q})&=&\Lambda({\bf
  k}_1,z_1,{\bf k}_2,z_2;{\bf q}) + \frac 1N \sum_{{\bf q}''} \Lambda({\bf
  k}_1,z_1,{\bf k}_2,z_2;{\bf q}'')\nonumber \\ && \hspace*{-20pt}
  \times G({\bf k}_1+{\bf q}'',z_1) )G({\bf k}_2+{\bf q}'',z_2)
  \Gamma({\bf k}_1+{\bf q}'',z_1,{\bf k}_2+{\bf q}'',z_2;{\bf q}-{\bf
  q}'')\, .
\end{eqnarray}

The one and two-particle irreducible functions, i.~e., self-energy
$\Sigma({\bf k},z)$ and vertex $\Lambda({\bf k}_1,z_1,{\bf k}_2,z_2;{\bf
  q})$ are the quantities that we have to approximate in order to determine
the one and two-particle characteristics of a disordered system. The two
functions are not completely independent.  In a conserving and
thermodynamically consistent approximation we have a generalized
differential Ward identity\cite{Baym61}
\begin{eqnarray}
  \label{eq:Ward_differential}
  \Lambda({\bf k}_1,z_1,{\bf k}_2,z_2;{\bf q})&=&
  \frac{\delta\Sigma({\bf k}_1,z_1,{\bf k}_2,z_2;U)}{\delta G({\bf
  k}_1+{\bf q},z_1,{\bf k}_2+{\bf q},z_2;U)}\bigg|_{U=0} \, .
\end{eqnarray}
We could use it for the determination of the irreducible vertex if we knew
explicitly the self-energy as a functional of the averaged propagator in
the presence of the external disturbance $U$.  It is rarely the case. We,
however, show in Sec.~\ref{sec:Ward_identity} how to use an integral form
of the Ward identity to determine the self-energy from the known
irreducible vertex $\Lambda$. It is then sufficient to construct an
approximation for the two-particle irreducible vertex $\Lambda$, which will
be done in the following subsections.

\subsection{Local approximation}
\label{sec:local}

We start building approximations to the two-particle vertex function
from a local solution where we completely loose momentum dependence.
The local approximation means that we use only site-diagonal
one-particle propagators in the perturbation diagrammatic expansion
for the irreducible functions. The local approximation is best derived
within the Baym-Kadanoff renormalized perturbation expansion in the
limit of high spatial dimensions $d\to\infty$.\cite{Janis99a} In this
limit we have the following asymptotics for the one-particle functions
\begin{equation}
  \label{eq:1P_separation}
  G=G^{diag}[d^0]+G^{off}[d^{-1/2}]\, ,\qquad \Sigma=\Sigma^{diag}[d^0]
  +\Sigma^{off}[d^{-3/2}]
\end{equation}
leading to separation of the diagonal (local) and off-diagonal (nonlocal)
parts. In the strict limit $d=\infty$ we can completely neglect the
off-diagonal elements and recover the CPA for the self-energy. The defining
equation in the presence of the external local disturbance $U$ reads
\begin{eqnarray}\label{eq:CPA_equation}
  \widehat{G}(z_1,z_2;U)&=&\left\langle \left[\widehat{G}^{-1}(z_1,z_2;U)
      +\widehat{\Sigma}(z_1,z_2;U) -\widehat{V}_i\right]^{-1}\right
  \rangle_{av}
\end{eqnarray}
where $\widehat{G}(z_1,z_2;U)=N^{-2}\sum_{{\bf k}_1{\bf
    k}_2}\widehat{G}({\bf k}_1,z_1,{\bf k}_2,z_2;U)$ is the local element
of the matrix one-particle propagator. The matrix character is forced by
the external disturbance which mixes different complex energies. Since we
are interested only in averaged two-particle functions in equilibrium, we
can resort to two energies and a two-by-two matrix
\begin{eqnarray}
  \label{eq:1P_matrix}
  \widehat{G}^{-1}({\bf k}_1,z_1,{\bf k}_2,z_2;U)&=&
  \begin{pmatrix}z_1-\epsilon({\bf k}_1) -\Sigma(z_1) & U -
    \Sigma(z_1,z_2;U)\\
    U - \Sigma(z_2,z_1;U) & z_2-\epsilon({\bf k}_2) -\Sigma(z_2)
    \end{pmatrix}
\end{eqnarray}
where $\epsilon({\bf k})$ is the lattice dispersion relation.

We use the Ward identity (\ref{eq:Ward_differential}) to determine the
two-particle irreducible vertex in equilibrium
\begin{eqnarray}\label{eq:2IP_vertex}
  \Lambda(z_1,z_2)&=&\frac{\delta\Sigma_U(z_1,z_2)}{\delta
    G_U(z_1,z_2)}\bigg|_{U=0} \nonumber \\ &=&\frac 1{G(z_1)G(z_2)}\left[1-
    {\left\langle \frac 1{ 1+\left(\Sigma(z_1)-V_i\right)G(z_1)}\frac 1{
      1+\left( \Sigma(z_2)-V_i\right)G(z_2)}\right\rangle_{av}}^{-1}
\right] \, .
\end{eqnarray}
The full vertex function is then determined from the Bethe-Salpeter
equation (\ref{eq:eh_BS}) where the one-particle propagators are replaced
with the local ones. The integral equation reduces to an algebraic one and
we obtain an explicit representation
\begin{eqnarray}\label{eq:local_vertex}
  \gamma(z_1,z_2)&=&\frac{\Lambda(z_1,z_2)}{1-\Lambda(z_1,z_2)G(z_1)G(z_2)}
  \, .
\end{eqnarray}
For later convenience we denoted the full local two-particle vertex with
the lower-case $\gamma$.

Note that the above local approximation coincides with the CPA only for the
self-energy and the local part of the vertex function. The CPA two-particle
vertex is obtained from Eq.~\eqref{eq:eh_BS} where only the irreducible
vertex $\Lambda(z_1,z_2)$ is assumed local, but not the one-particle
propagators.\cite{Velicky69}  Here we deliberately separated the
contribution from the off-diagonal propagator.  We describe it within the
parquet approach as a correction to the local approximation to the vertex
function.

\subsection{Nonlocal contributions: Parquet approach}
\label{sec:parquet}
 
To go beyond the local approximation we have to distinguish two
one-particle propagators. First, we have the propagator from the local
approximation that we denote $G^{loc}({\bf k},z)$, from which we
actually need only the local element $G^{loc}(z)=N^{-1}\sum_{\bf
  k}G^{loc}({\bf k},z)$. This propagator is defined by the Dyson
equation with the local self-energy $\Sigma(z)$ from
Eq.~\eqref{eq:CPA_equation} with $U=0$.  Second, we have to introduce
a new propagator $G({\bf k},z)$ that is defined by the Dyson equation
with a nonlocal self-energy $\Sigma({\bf k},z)$ to be determined
later.  It is treated in the equations for the two-particle vertex as
an external function with appropriate analytic properties.  In the
expansion beyond the local approximation we use a perturbed propagator
$\widetilde{G}({\bf k},z)=G({\bf k},z)-G^{loc}(z)$ to avoid multiple
summations of single-site diagrams contained in the local
approximation.

We classify {\em nonlocal} contributions to the two-particle vertex by the
type of the correlated two-particle propagation. We either simultaneously
propagate an electron and a hole or two electrons (holes). Diagrammatically
it means that we connect {\em spatially distinct} two-particle scattering
events with antiparallel or parallel pairs of one-particle propagators.
Multiple scatterings of pairs of the same type define a channel of
two-particle irreducibility. We call a diagram two-particle irreducible if
it cannot be split into separate parts by cutting simultaneously either
electron-hole or electron-electron propagators. The two definitions of the
two-particle irreducibility lead to topologically inequivalent irreducible
functions and to different Bethe-Salpeter equations for the full vertex. In
each Bethe-Salpeter equation the two-particle functions are interconnected
via one-particle propagators in a different manner. We can generically
represent the Bethe-Salpeter equations as
\begin{eqnarray} \label{eq:2P-reducible}
\Gamma({\bf k}_1,z_1,{\bf k}_2,z_2;{\bf q})&=&\Lambda^\alpha({\bf
k}_1,z_1,{\bf k}_2,z_2;{\bf q}) +\left[\Lambda^\alpha
\widetilde{G}\widetilde{G} \odot \Gamma \right]({\bf k}_1,z_1,{\bf
k}_2,z_2;{\bf q}) .
\end{eqnarray}
We used $\odot$ for the channel-dependent multiplication of the
two-particle functions. Here $\Lambda^\alpha$ is the irreducible vertex in
the $\alpha$-channel. Although the irreducible functions are different in
different channels the solution of the Bethe-Salpeter equations must always
be the same, the full two-particle vertex $\Gamma$. The matrix
multiplication in momentum space in the electron-hole and electron-electron
channels, respectively explicitly reads
\begin{subequations}\label{eq:conv}
\begin{eqnarray}\label{eq:conv_eh} &&
\left[\widehat{X}GG\bullet\widehat{Y}\right]({\bf k}_1,z_1,{\bf
k}_2, z_2;{\bf q})=\frac 1N\sum_{{\bf q}''} X({\bf k}_1,z_1,{\bf
k}_2,z_2;{\bf q}'') \nonumber \\&&\hspace*{40pt} \times G({\bf
k}_1+{\bf q}'',z_1)G({\bf k}_2+{\bf q}'',z_2) Y({\bf k}_1+{\bf
q}'',z_1,{\bf k}_2+{\bf q}'',z_2;{\bf q}-{\bf q}''), \\
\label{eq:conv_ee} && 
\left[\widehat{X}GG\circ\widehat{Y}\right]({\bf k}_1,z_1,{\bf
k}_2,z_2;{\bf q})= \frac 1N\sum_{{\bf q}''} X({\bf k}_1,z_1,{\bf
k}_2+{\bf q}'',z_2;{\bf q}-{\bf q}'') \nonumber \\ 
&&\hspace*{40pt} \times G({\bf k}_1 +{\bf q}-{\bf q}'',z_1)
G({\bf k}_2+{\bf q}'',z_2) Y({\bf k}_1+{\bf q}-{\bf
q}'',z_1,{\bf k}_2,z_2;{\bf q}'')\, . \end{eqnarray}

Electron-hole and electron-electron channels are not the only inequivalent
representations of multiple two-particle scatterings. If we allow for
hopping between distant sites we no longer can distinguish between multiple
scatterings of distinct electron-hole pairs or nonlocal self-scatterings of
a single particle, i.~e., scatterings between the incoming and the outgoing
particle at the two-particle vertex. We hence have to introduce a third
two-particle irreducibility and Bethe-Salpeter equation.  We call this
third type of two-particle scatterings ``vertical channel'', since we
renormalize only one line of the pair and the ladder of multiple
scatterings grows ``vertically'' above (below) the two-particle vertex. We
can write the third multiplication scheme for two-particle quantities as
follows
\begin{eqnarray}
  \label{eq:conv_U}
 && \left[\widehat{X}GG\star\widehat{Y}\right]({\bf k}_1,z_1,{\bf
   k}_2,z_2;{\bf q})= \frac 1N\sum_{{\bf k}''}X({\bf k}_1,z_1,{\bf
   k}_1+{\bf q},z_1;{\bf k}''-{\bf k}_1) \nonumber \\ &&\hspace*{100pt}
  \times G({\bf k}'',z_1)G({\bf k}''+{\bf q},z_1) Y({\bf k}'',z_1,{\bf
    k}_2,z_2;{\bf q}) .
\end{eqnarray}
\end{subequations}
In this case only the particle with energy $z_1$ is renormalized.
Analogously we introduce self-scattering vertex corrections to the particle
with energy $z_2$. The vertical channel actually splits into two, upper and
lower parts according to whether we renormalize the first or the second
energy in the pair.

Using the above multiplication schemes for different two-particle channels
we can write down the corresponding Bethe-Salpeter equations explicitly.
We choose a subscript $\alpha=\pm $ for complex half-planes from which we
take the particle energy. We use the standard diagrammatic representation
for these equations and obtain for the electron-hole and electron-electron
channels
\begin{subequations}\label{eq:BS_figures}
\begin{eqnarray}\label{eq:BS_figure_eh}
&&\hspace*{-10pt}\epsfig{figure=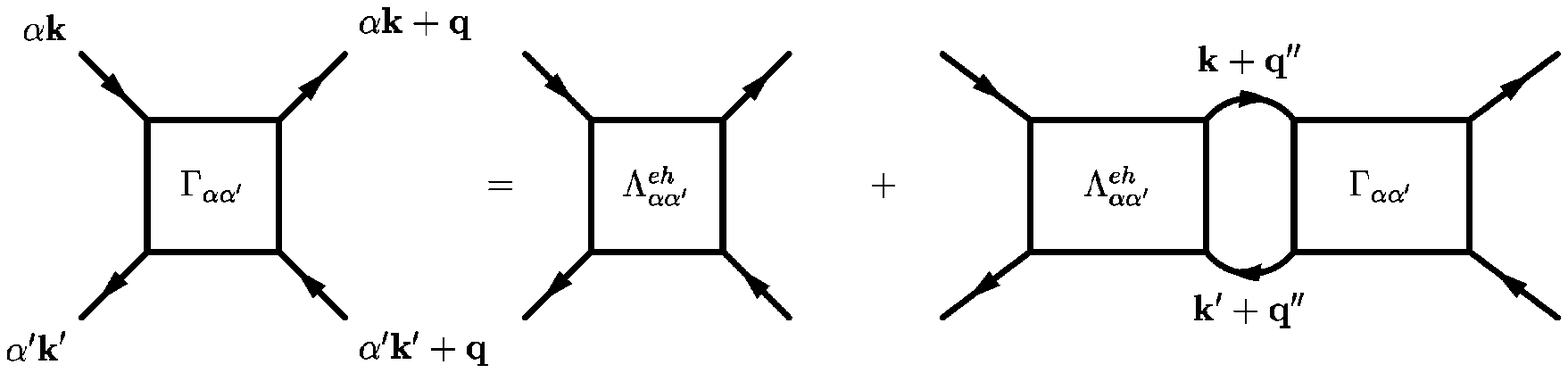,  
  height=3.5cm}\nonumber \\
\end{eqnarray}
\begin{eqnarray}
 \label{eq:BS_figure_ee} 
 &&\hspace*{-0pt}
\epsfig{figure=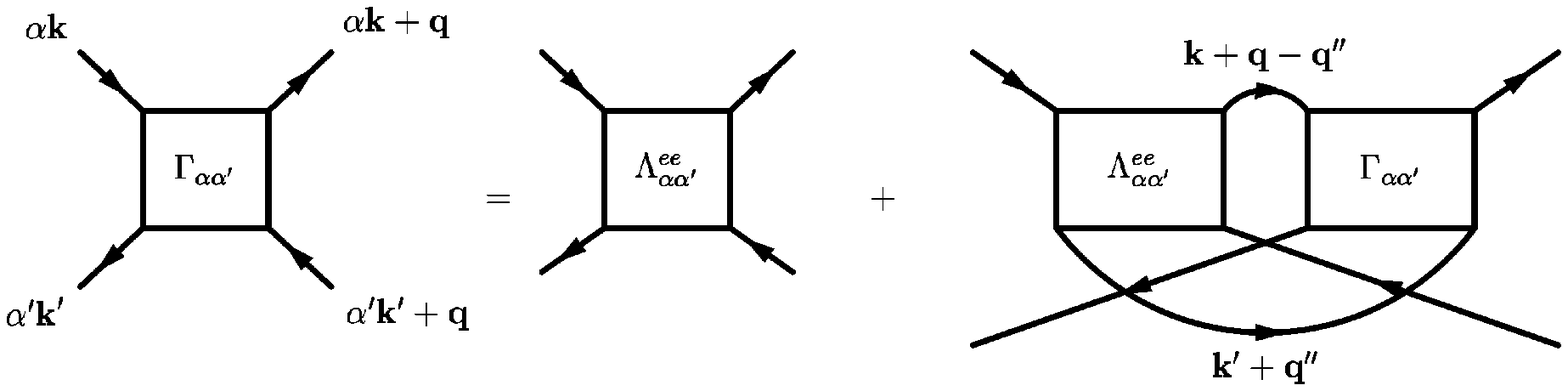,
  height=3.9cm}\nonumber \\
\end{eqnarray}
The Bethe-Salpeter equation in the vertical channel is split into two.
First we account for self-scattering vertex corrections to the upper line
and then the same for the lower one. We then have
\begin{eqnarray} \label{eq:BS_figure_u1}
   &&\hspace*{-30pt}
   \epsfig{figure=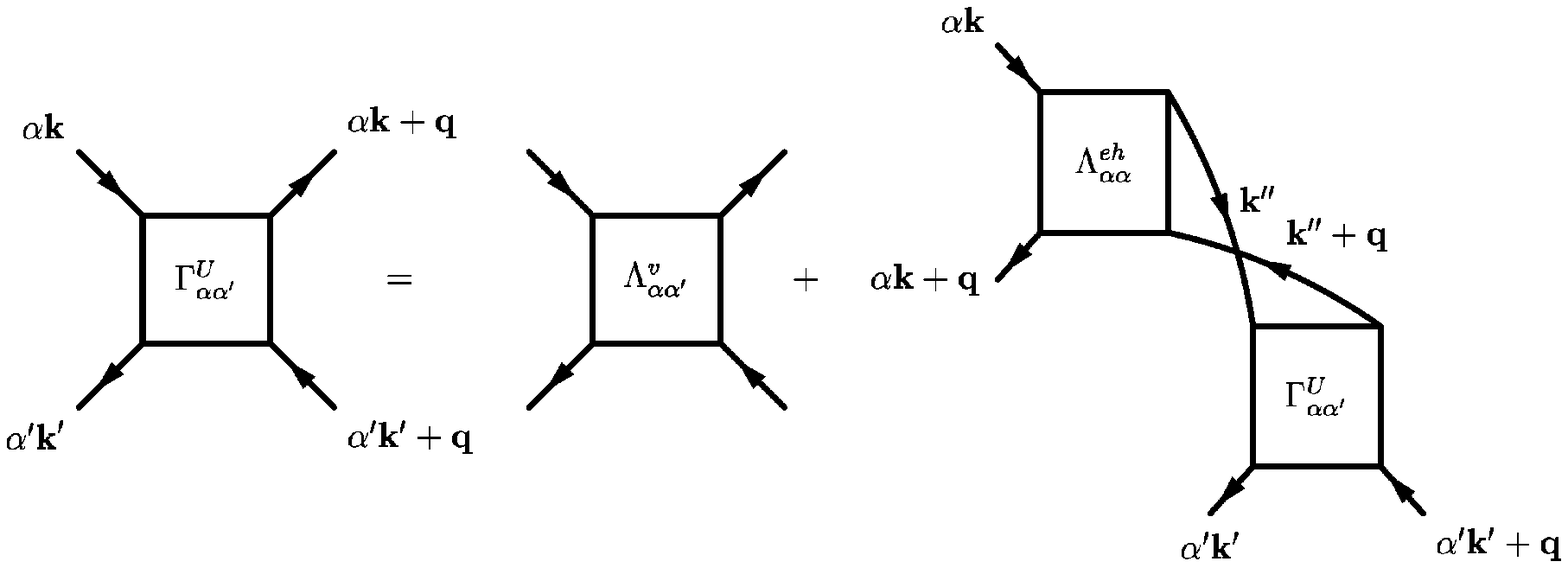,height=5cm}
\end{eqnarray}
\begin{eqnarray}
   \label{eq:BS_figure_u2}
     &&\hspace*{-30pt}
   \epsfig{figure=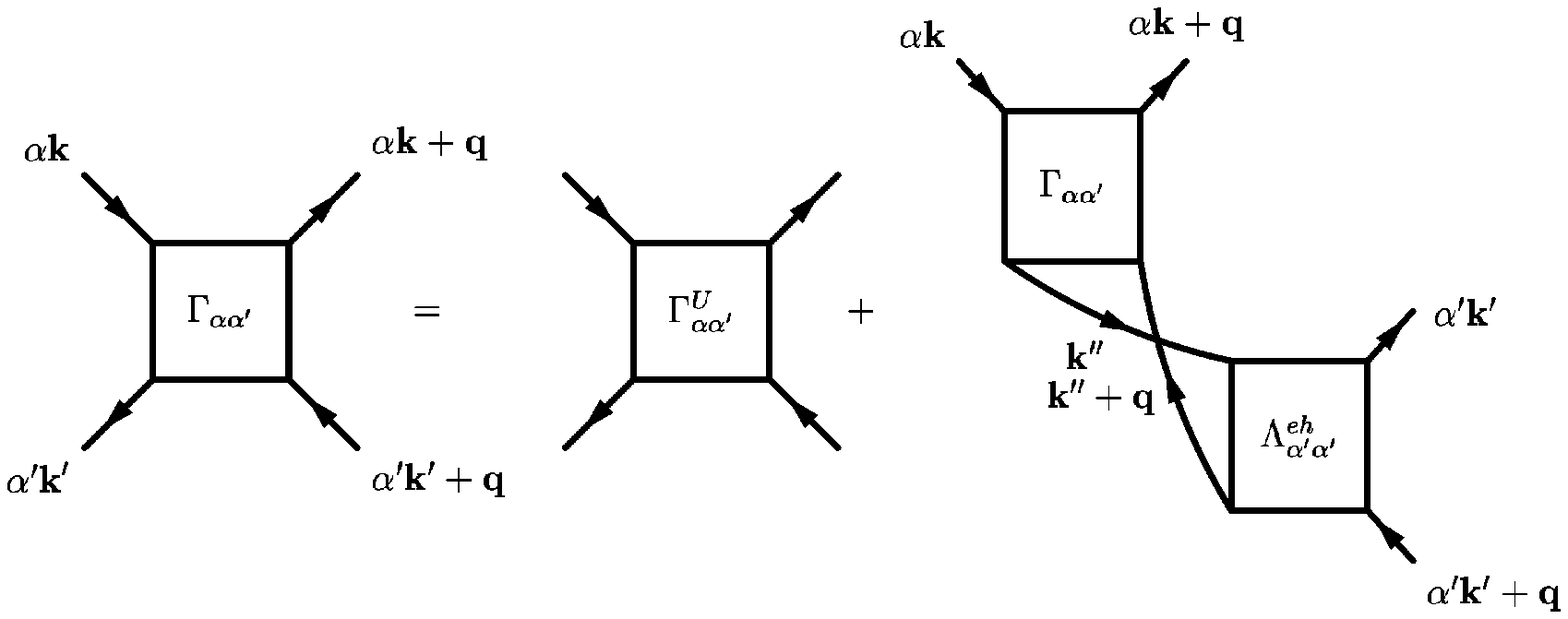,height=5cm}
\end{eqnarray}
\end{subequations}
Note that it is the irreducible vertex from the electron-hole channel
$\Lambda^{eh}$ that determines the kernel of the integral equations
\eqref{eq:BS_figure_u1} and \eqref{eq:BS_figure_u2}. The new vertex
$\Lambda^v$ enters only via the absolute term.  The irreducible vertices
$\Lambda^v$ and $\Lambda^{eh}$ are not generally equal and it is useful for
our construction to distinguish the vertical channel from the other two.

In all these equations the double-prime momenta are summed over. The full
lines stand for the perturbed propagators $\widetilde{G}({\bf k},z)$.
Otherwise we would encounter multiple summations of single-site
scatterings. Separation of the local and nonlocal contributions is
important, since in the local approximation the three channels coincide.
This can be seen from Eq.~\eqref{eq:conv} when we insert a local
propagator.  This fact is physically obvious, since we cannot distinguish
between the particle and the hole in single-site multiple scatterings. The
electron and the hole propagators equal. There is only one two-particle
irreducible vertex $\Lambda(z_1,z_2)$ in the local approximation (CPA).

We now use the topological inequivalence of the three channels. This means
that a reducible function from one channel is irreducible in the other
ones. If we denote $I$ the completely irreducible two-particle
vertex\cite{Note2a} we can write 
\begin{subequations}\label{eq:2P_irreducible}
\begin{eqnarray}
  \label{eq:2P_irreducible_sum}
  \Lambda^\alpha({\bf k}_1,z_1,{\bf k}_2,z_2;{\bf q}) &= &I({\bf
    k}_1,z_1,{\bf k}_2,z_2;{\bf q})
  +\sum_{\alpha'\neq\alpha} \left(\Gamma({\bf k}_1,z_1,{\bf k}_2,z_2;{\bf
      q}) - \Lambda^{\alpha'}({\bf k}_1,z_1,{\bf k}_2,z_2;{\bf q})
    \right) \, ,
\end{eqnarray}
since the reducible function is a difference between the full and the
irreducible vertex from a given channel. We use the Bethe-Salpeter
equations \eqref{eq:BS_figures} on the r.h.s. of
Eq.~(\ref{eq:2P_irreducible_sum}) from the corresponding $\alpha'$ channel
to get rid of the full vertex. We then obtain a set of equations for the
channel-dependent irreducible vertices $\Lambda^\alpha$. The completely
irreducible vertex is input to these so-called parquet equations. The
lowest-order contribution to the input function is the local two-particle
vertex $\gamma$.  The parquet equations in a generic form are
\begin{eqnarray}
  \label{eq:2P_irreducible_parquets}
  \Lambda^\alpha({\bf k}_1,z_1,{\bf k}_2,z_2;{\bf q}) &=&  \gamma(z_1,z_2)
  +\sum_{\alpha'\neq\alpha}\left\{1-\left[\Lambda^{\alpha'}\widetilde{G}
      \widetilde{G}\right] \odot\right\}^{-1}
  \left[\Lambda^{\alpha'}\widetilde{G} \widetilde{G}\odot
    \Lambda^{\alpha'}\right]({\bf k}_1,z_1,{\bf k}_2,z_2;{\bf q})\, .
\end{eqnarray}
\end{subequations}
Special attention should be paid to the vertical channel, $\alpha'=v$,
where we have a two-step construction and Bethe-Salpeter equation
\eqref{eq:BS_figure_u1} and \eqref{eq:BS_figure_u2}. The second term on the
r.h.s. of Eq.~\eqref{eq:2P_irreducible_parquets} should be replaced by
$\left\{1-\left[\Lambda^{eh}\widetilde{G}\widetilde{G}\right]
  \star\right\}^{-1}\Lambda^v \left\{1-\star\left[\widetilde{G}
    \widetilde{G} \Lambda^{eh}\right]\right\}^{-1} -\Lambda^v$.

Equations \eqref{eq:2P_irreducible} constitute the parquet approximation
for the irreducible vertices for given one-particle propagators $G({\bf
  k},z)$, $G^{loc}(z)$ and the local two-particle vertex $\gamma(z_1,z_2)$,
that again is a function of $G^{loc}$ from Eq.~\eqref{eq:local_vertex}.
The latter propagator is determined from the local approximation, CPA. The
former is treated here as a function of a self-energy $\Sigma({\bf k},z)$
which we connect to the vertex functions later on. Parquet equations
\eqref{eq:2P_irreducible_parquets} form a set of nonlinear integral
equations self-consistently determining the two-particle irreducible vertex
functions $\Lambda^\alpha[\Sigma;G^{loc}]$.

The parquet equations represent a substantial extension of the local
approximation. They are, however, much more difficult to solve than cluster
or other short-range extensions of the CPA at the one-particle level. To
get a feeling how a solution to the parquet equations looks like we resort
to the asymptotic limit $d\to\infty$ where the corrections to the local
vertex $\gamma(z_1,z_2)$ in Eq.~(\ref{eq:2P_irreducible_parquets})
asymptotically vanish.\cite{Janis99a} The two-particle self-consistence
becomes asymptotically insignificant and such a situation can be dealt with
exactly.

\subsection{Asymptotic limit of high spatial dimensions}
\label{sec:vertex_high_dim}

We know that the off-diagonal, nonlocal elements are scaled to zero in high
dimensions and loose their weight with respect to the local
ones.\cite{Metzner89} However, when summed over the lattice sites they can
produce finite contributions even in the limit $d=\infty$. It is the case
of the two-particle vertex. In the leading asymptotic order the irreducible
vertices in the Bethe-Salpeter equations become local and coincide with
$\gamma(z_1,z_2)$. The asymptotic behavior of the full vertex then depends
upon which matrix element we calculate. We obtain different asymptotic
solutions in different channels that we denote $\Gamma^{\alpha}$
\cite{Janis99a}. The corresponding asymptotic Bethe-Salpeter equations in
the three channels have the following diagrammatic representation
\begin{subequations}\label{eq:BS_infty}
\begin{eqnarray}
&&\hspace*{-10pt}\epsfig{figure=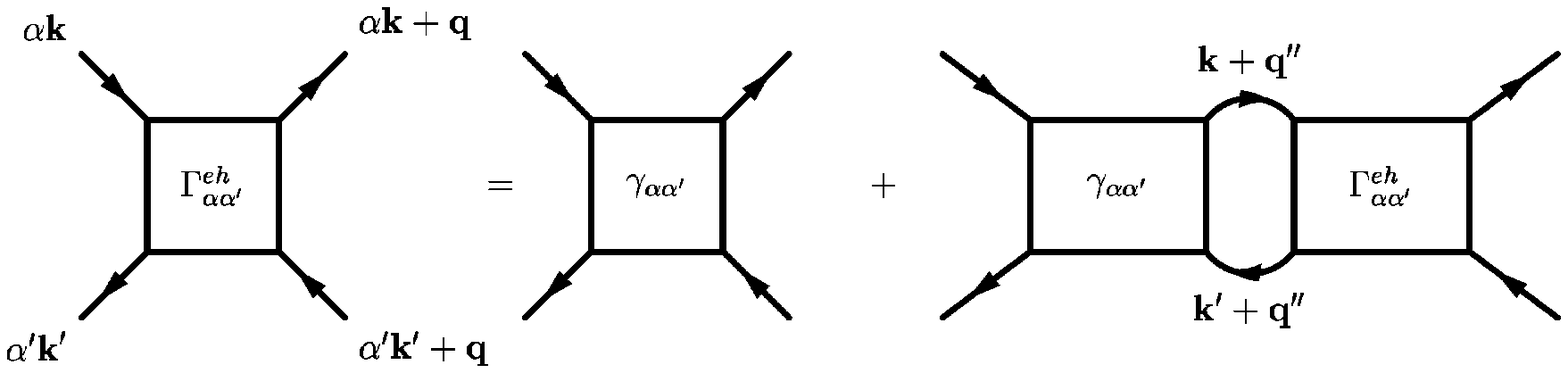,
  height=3.5cm}\nonumber \\
\end{eqnarray}
\begin{eqnarray}
 &&\hspace*{-0pt}
\epsfig{figure=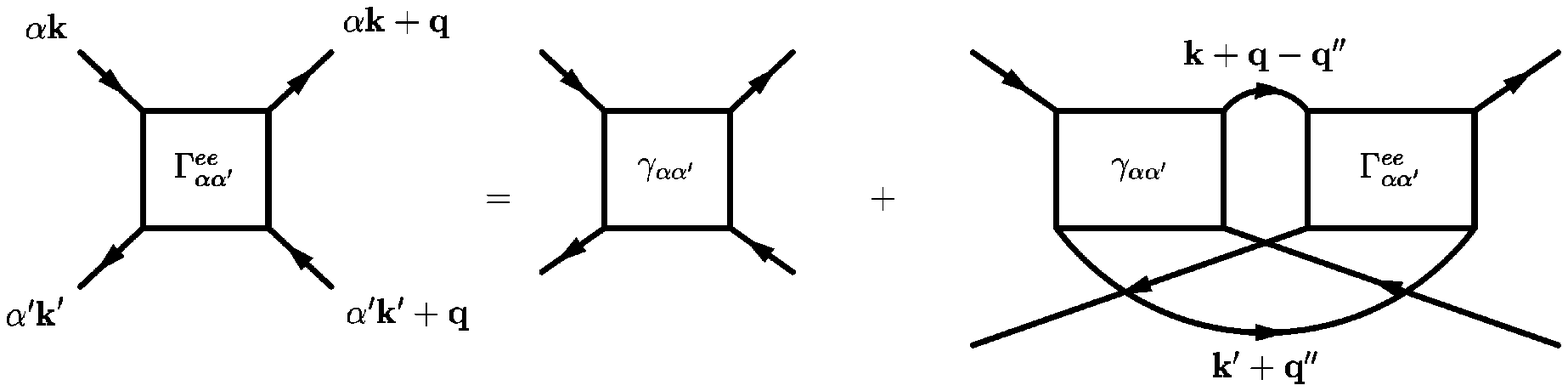,
  height=3.9cm}\nonumber \\
\end{eqnarray}
\begin{eqnarray}
   &&\hspace*{-30pt}
   \epsfig{figure=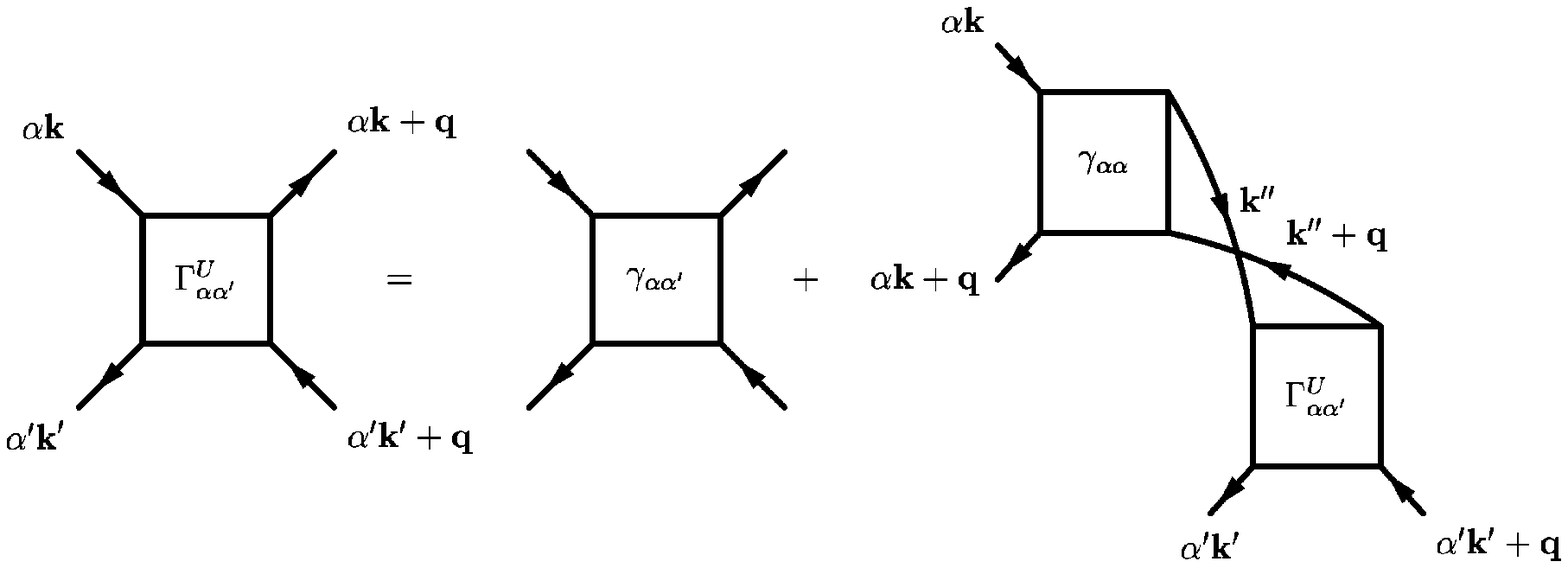,height=5cm}
\end{eqnarray}
\begin{eqnarray}
      &&\hspace*{-30pt}
\epsfig{figure=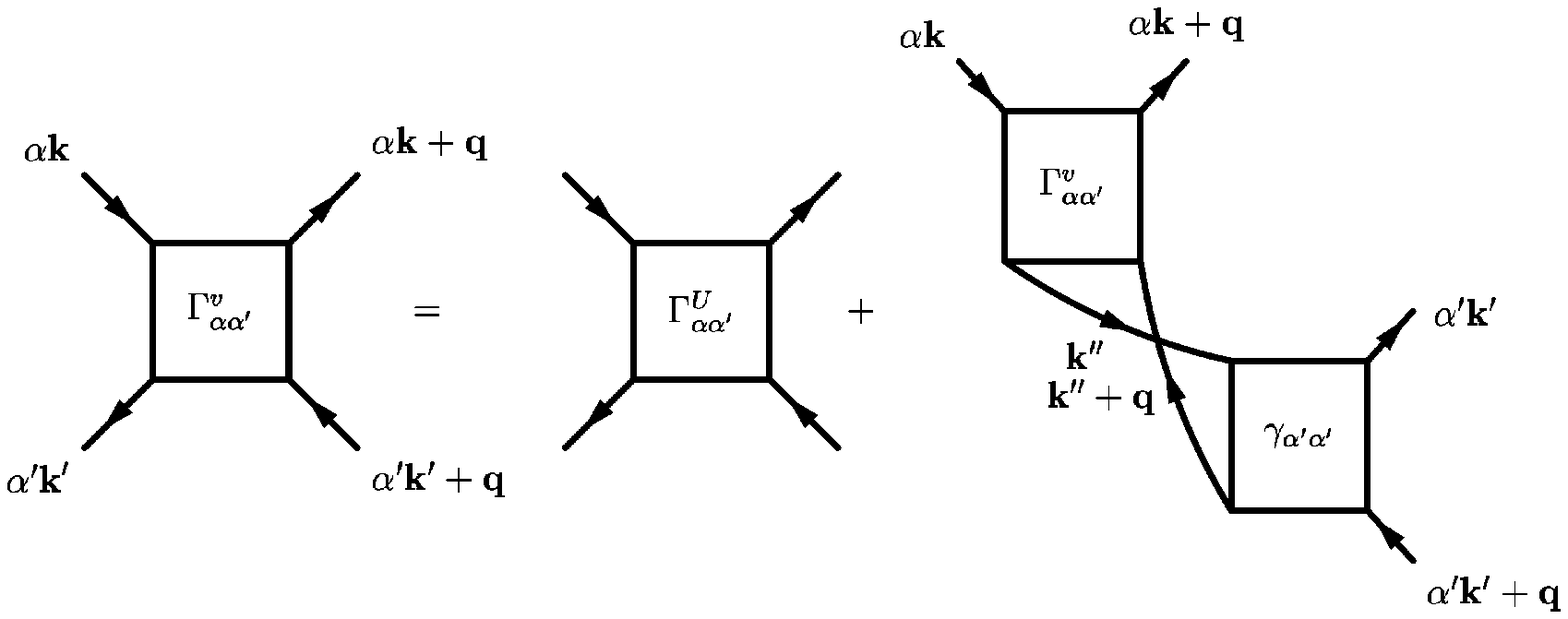,height=5cm}
\end{eqnarray}
\end{subequations}
The one-particle propagator in the high-dimensional Bethe-Salpeter
equations equals the CPA solution, $G({\bf k},z)=G^{loc}({\bf k},z)$, or
$\Sigma({\bf k},z)=\Sigma^{CPA}(z)$.  We sum over the intermediate momenta
in the Bethe-Salpeter equations and also the nonlocal part of $G^{loc}({\bf
  k},z)$ contributes to the leading asymptotic behavior of the solution for
fixed two-particle momenta.

The Bethe-Salpeter equations (\ref{eq:BS_infty}) in high dimensions become
algebraic and can be solved in closed form. If we denote
\begin{equation}
  \label{eq:off-diagonal_bubble}
  Y^{\pm}({\bf q};z_1,z_2)=\sum_{\bf k} G^{off}({\bf k},z_1) G^{off}({\bf
    q}\pm {\bf k},z_2)
\end{equation}
the solutions in the three channels are
\begin{subequations}  \label{eq:infty_channels}
\begin{eqnarray}
  \label{eq:infty_eh_channel}
  \Gamma^{eh}({\bf q};z_1,z_2)&=& \frac{\gamma(z_1,z_2)}
  {1-\gamma(z_1,z_2)Y^+({\bf q};z_1,z_2) }  \\ \label{eq:infty_ee_channel}
  \Gamma^{ee}({\bf q};z_1,z_2)&=& \frac{\gamma(z_1,z_2)}
  {1-\gamma(z_1,z_2)Y^-({\bf q};z_1,z_2) }  \\ \label{eq:infty_v_channel}
    \Gamma^{v}({\bf q};z_1,z_2)&=& \gamma(z_1,z_2) \prod_{i=1}^2 \frac{1}
  {\left[1-\gamma(z_i,z_i)Y^+({\bf q};z_i,z_i) \right]}\ .
\end{eqnarray}
\end{subequations}

The three different vertex functions represent the three asymptotic
solutions we obtain in the limit of high spatial dimensions. Each solution
is characterized by a two-particle momentum that is kept fixed. They
are in the notation from the preceding subsection %
${\bf k}_1-{\bf k}_2$, ${\bf k}_1+ {\bf k}_2 +{\bf q}$, and ${\bf q}$ for
the electron-hole, electron-electron, and the vertical channel,
respectively. The full vertex $\Gamma$ in high dimensions reduces to a sum
\begin{eqnarray}\label{eq:d_infty_vertex}
 \Gamma({\bf k}_1,z_1,{\bf k}_2,z_2; {\bf q})&=&
 \Gamma^{eh}({\bf k}_1-{\bf  k}_2;z_1,z_2)+ \Gamma^{ee}({\bf k}_1+{\bf
   k}_2+{\bf q};z_1,z_2) +\Gamma^v({\bf q};z_1,z_2) 
-2\gamma(z_1,z_2)
\end{eqnarray}
where we had to subtract appropriately the local vertex to avoid double
counting. Note that the CPA vertex as derived by Velick\'y\cite{Velicky69}
equals $\Gamma^{eh}$ and does not correspond to the leading
high-dimensional asymptotics of the exact two-particle vertex.

Representation (\ref{eq:infty_channels}) uses the natural high-dimensional
separation of the diagonal and off-diagonal parts of the one-particle
propagator.  We can rewrite the asymptotic solution
\eqref{eq:infty_channels} to another form with full one-particle
propagators, which is more appropriate for finite-dimensional systems. We
denote a two-particle bubble
\begin{eqnarray}\label{eq:bubble}
  \chi^{\pm}({\bf q};z_1,z_2)&=&\frac 1N\sum_{\bf k} G({\bf k},z_1)G({\bf
    q}\pm{\bf k},z_2)  \, .
\end{eqnarray}
If we further use relation (\ref{eq:local_vertex}) we obtain a new
representation for the two-particle vertex
\begin{eqnarray}\label{eq:2P_vertex_hd}
  \Gamma({\bf k}_1,z_1,{\bf k}_2,z_2;{\bf q})&=&\Lambda(z_1,z_2)\left\{\frac
    1{1-\Lambda(z_1,z_2)\chi^{+}({\bf k}_2-{\bf k}_1;z_1,z_2)}
    -\frac 2{1-\Lambda(z_1,z_2)G(z_1)G(z_2)}
    \right.\nonumber\\[10pt] && \left.\hspace*{-60pt}
    +\ \frac{\displaystyle \prod_{i=1}^2
    \frac{1-\Lambda(z_i,z_i)G(z_i)G(z_i)}
  {\left[1-\Lambda(z_i,z_i)\chi^+({\bf q};z_i,z_i)\right]}}
{1-\Lambda(z_1,z_2)G(z_1)G(z_2)}  + \frac1{1-\Lambda(z_1,z_2)\chi^{-} ({\bf
    k}_1+{\bf k}_2- {\bf q};z_1,z_2)}  \right\}  \,
\end{eqnarray}
where $\Lambda(z_1,z_2)$ is the CPA irreducible vertex.

Note that the asymptotic solution for the two-particle vertex can no longer
be represented via a single Bethe-Salpeter equation. It is because we have
three asymptotically equally important but topologically different
contributions. Each of them is marked by a different fixed two-particle
momentum.

\section{Ward identities and the self-energy}
\label{sec:Ward_identity}

In the preceding section we derived a closed set of parquet equations
determining the irreducible two-particle vertex functions from the given
propagators $G^{loc}(z)$ and $G({\bf k},z)$. The former is the
site-diagonal part of the CPA propagator but the latter has not yet been
specified. It is determined by a self-energy $\Sigma({\bf k},z)$ from the
Dyson equation \eqref{eq:av_1PP}. We have not yet demanded any relation
between the self-energy and the vertex functions.  We, however, know that
thermodynamic consistence demands that the irreducible one and two-particle
functions be not independent. They are related by the differential Ward
identity \eqref{eq:Ward_differential}. To turn the parquet equations a
conserving approximation we have to fulfill the Ward identity and relate
the self-energy to the solution of the parquet equations.  The functional
differential identity \eqref{eq:Ward_differential} is of little practical
help in calculating the self-energy from the irreducible two-particle
functions.  Fortunately there are integral forms of the Ward identity
relating one and two-particle averaged functions that can be used to
determine the self-energy from a known two-particle irreducible vertex.

A first integral Ward identity was derived by Velick\'y\cite{Velicky69} in
the framework of the coherent-potential approximation. It holds quite
generally beyond the CPA and reads in our notation
\begin{subequations}\label{eq:W_BV}
\begin{equation}
\label{eq:W_BV_lattice}
\sum_j G^{(2)}_{ij,jk}(z_1,z_2)=-\frac 1{\Delta z} \Delta G_{ik}\, ,
\end{equation}
where $\Delta z=z_1-z_2$, $\Delta G_{ij}=G_{ij}(z_1)-G_{ij}(z_2)$. We show
in Appendix~A that this identity is a consequence of completeness of the
eigenstates of the Hamiltonian and hence of probability conservation.  This
identity can be rewritten in momentum space
\begin{eqnarray}
  \label{eq:W_BV_momentum}
  \frac 1N\sum_{\bf q}G^{(2)}({\bf k},z_1,{\bf k},z_2;{\bf q})&=& - \frac
  1{\Delta z} \left[ G({\bf k},z_1)-G({\bf k},z_2)\right]\, .
\end{eqnarray}
\end{subequations}

Ward identity \eqref{eq:W_BV} is inconvenient for application, since it
connects one and two-particle averaged functions. We better had a relation
between irreducible one and two-particle functions alike
Eq.~\eqref{eq:Ward_differential}. An integral form of such a Ward identity
was proven diagrammatically by Vollhardt and W\"olfle\cite{Vollhardt80}
and reads 
\begin{equation}
\label{eq:W_VW}
\Sigma({\bf k}_1,z_1) -\Sigma({\bf k}_2,z_2) = \frac
1{N}\sum_{{\bf q}}\Lambda^{eh}({\bf k}_1,z_1,{\bf k}_2,z_2;{\bf q})
\left[G({\bf k}_1+{\bf q},z_1) - G({\bf k}_2+{\bf q},z_2)\right]\, .
\end{equation}
The diagrammatic derivation of the Vollhardt-W\"olfle identity
\eqref{eq:W_VW} utilizes a symmetry of the Anderson disordered model but
does not establish a direct relation to conservation laws as the derivation
of the Velick\'y identity \eqref{eq:W_BV}. We show in Appendix~A that the
Vollhardt-W\"olfle identity with ${\bf k}_1={\bf k}_2$ follows directly
from Eq.~\eqref{eq:W_BV} and the Bethe-Salpeter equation. At least this
simplified form of the Vollhardt-W\"olfle identity can be shown to be a
consequence of probability conservation.\cite{Note3}

To complete the parquet approximation for the two-particle irreducible
vertex functions we can use Eq.~\eqref{eq:W_VW} for ${\bf k}_1={\bf k}_2$ and
limiting values of the complex energies along the real axis. We chose
$z_1=\omega_+$ and $z_2=\omega_-$ where $\omega_\pm=\omega\pm i\eta$ and
$\eta\searrow0$. We obtain a relation
\begin{subequations}\label{eq:SE_vertex}
\begin{eqnarray}
  \label{eq:SE_vertex_im}
  \mbox{Im}\Sigma({\bf k},\omega_+)=\frac 1N\sum_{{\bf k}'}\Lambda^{eh}
  ({\bf k},\omega_+,{\bf k},\omega_-;{\bf k}-{\bf k}')\mbox{Im} G({\bf
  k}',\omega_+)
\end{eqnarray}
determining the imaginary part of the self-energy along the real axis from
the known irreducible vertex in the electron-hole channel. A similar
formula can be derived for the irreducible vertex from the
electron-electron channel, cf. Appendix~A.

Next we rely on analytic properties of the self-energy in the upper and
lower complex half-planes and determine its real part along the real axis
from the Kramers-Kronig relation
\begin{eqnarray}
  \label{eq:SE_vertex_re}
  \mbox{Re}\Sigma({\bf k},\omega_+)&=& P\int_{-\infty}^{\infty}
   \frac{d\omega'}{\pi} \ \frac {\mbox{Im}\Sigma({\bf
   k},\omega'_+)}{\omega'-\omega}\, .
\end{eqnarray}\end{subequations}
The self-energy from the cut along the real axis can be analytically
continued to energies in the upper and lower complex half-planes. The
one-particle perturbed propagator from the parquet equations
\eqref{eq:2P_irreducible_parquets} can now be defined from the irreducible
vertex via
\begin{eqnarray}
  \label{eq:1P_Dyson}
  \widetilde{G}({\bf k},\omega_+)&=&\left[\omega_+-\epsilon({\bf k})-\Sigma({\bf
      k},\omega_+) \right]^{-1} - G^{loc}(\omega_+)
\end{eqnarray}
with the self-energy from Eq.~\eqref{eq:SE_vertex} and the local propagator
from Eq.~\eqref{eq:CPA_equation}.

Equations \eqref{eq:SE_vertex}, \eqref{eq:1P_Dyson} complete the parquet
approach to disordered systems and make it a consistent and conserving
scheme approximating simultaneously both the one and two-particle
irreducible functions.  Neither the Bethe-Salpeter equations
\eqref{eq:2P_irreducible_parquets} nor Eqs.~\eqref{eq:SE_vertex} violate
analytic properties, unless there is a transition to another phase such as
Anderson localization.  The analyticity of the parquet approximation in the
diffusive regime is then completely determined by its input, the
two-particle local CPA vertex $\gamma(z_1,z_2)$ that is known to possess
Herglotz analyticity.\cite{Mueller-Hartmann73}
 
The advantage of the parquet approach to disordered systems is that we do
not need to bother about the diagrams contributing to the self-energy so
that to obtain a consistent approximation with the proper analytic behavior
of the averaged Green functions.\cite{Note4}  An explicit sufficient
condition for analyticity of a solution of the parquet approximation from
Eq.~\eqref{eq:SE_vertex_im} is
\begin{eqnarray}
  \label{eq:parquet_analyticity}
  \Lambda^{eh} ({\bf k},\omega_+,{\bf k},\omega_-;{\bf q})&\ge& 0\, .
\end{eqnarray}
It can be checked in each step of iterations toward the full solution of
the parquet approximation for the one and two-particle irreducible
functions.

\section{Electrical conductivity}
\label{sec:conductivity}

The CPA constitutes a rather good approximation for the one-particle
self-energy but completely fails to incorporate coherence in the
propagation of pairs of particles.  The emphasis in the parquet approach is
hence laid on a systematic construction of diagrammatic approximations for
nonlocal two-particle functions. In particular those determining transport
properties and reflecting Anderson localization. In this section we show
how a solution of the parquet approximation can be used in the calculation
of the electrical conductivity.

We use a Kubo formula for the electrical conductivity with the
current-current correlation function. If $\sigma_{\alpha\beta}$ is the
complex conductivity and $\Pi_{\alpha\beta}$ the current-current
correlation function we can write\cite{Mahan84}
\begin{eqnarray}\label{eq:conductivity}
  \sigma_{\alpha\beta}({\bf q},E_+)&=&\frac i{E}\Pi_{\alpha\beta}({\bf
    q},E_+)
\end{eqnarray}
where again $E_+=E+i0^+$. The current-current correlation function can be
expressed via a Kubo formula with the full two-particle vertex $\Gamma$
\begin{eqnarray}\label{eq:current_current}
 \Pi_{\alpha\beta}({\bf q},i\nu_m)&=&\frac{e^2}{4m^2}\frac 1{N^2}
 \sum_{{\bf  k},{\bf k}'} \left[\partial_\alpha\left(\epsilon({\bf k}+{\bf q})-
   \epsilon(-{\bf k})\right)\partial_\beta\left(\epsilon({\bf k}'+{\bf q})
   -\epsilon(-{\bf k}')\right)\right]\nonumber\\ && \hspace*{-50pt}
 \times k_B T\sum_{n=-\infty}^\infty \left\{G({\bf k},i\omega_n)G({\bf k}+{\bf q},
  i\omega_n+i\nu_m)\left[\delta({\bf k}-{\bf k}')\right.\right. \nonumber\\
 &&\left.\left.\hspace*{-40pt}+\ \Gamma\left({\bf k},i\omega_n, {\bf
         k}+{\bf q}, i(\omega_n+\nu_m); {\bf k}'-{\bf k})\right) G({\bf
       k}', i\omega_{n}) G({\bf k}'+{\bf q}, i\omega_{n}+i\nu_m)
   \right]\right\}\, .
\end{eqnarray}
Here $\omega_n=(2n+1)\pi T$ and $\nu_m=2m\pi T$ are Matsubara frequencies.

We are actually interested only in the real part of the complex
conductivity for real energies. We can then analytically continue the
expression on the r.h.s. of Eq.~\eqref{eq:current_current}. For the real
part of the conductivity we obtain
\begin{eqnarray}\label{eq:real_conductivity}
 && \mbox{Re}\ \sigma_{\alpha\beta}({\bf q},E)=-\frac{e^2}{4}\frac 1{N^2}\sum_{{\bf
      k},{\bf k}'} (v_\alpha({\bf k}+{\bf q})+v_\alpha({\bf k}))
  (v_\alpha({\bf k}'+{\bf q})+v_\beta({\bf k}'))\frac 1{2i}\sum_{\sigma}\sigma
  \frac 1{2i} \sum_{\tau}\tau
     \nonumber \\&&\hspace*{0pt} \int_{-\infty}^\infty\frac{d\omega}\pi
   \frac{f(\omega+E)-f(\omega)}{E}
\left\{G({\bf k},\omega+i\sigma 0^+) G({\bf k}+{\bf q},\omega+E+i\tau
     0^+)\left[\delta({\bf k}-{\bf k}') \right.\right. \nonumber\\&&
   \left.\left.\hspace*{0pt} +\Gamma({\bf k},\omega+i\sigma 0^+,{\bf k}
       +{\bf q}, \omega+E+i\tau 0^+; {\bf k}'-{\bf k})    G({\bf k}',\omega
       +i\sigma 0^+) G({\bf k}',\omega+i\tau 0^+)\right]\right\}
\end{eqnarray}
where $f(E)$ is the Fermi function and we denoted the group velocity
$v_\alpha({\bf k})=m^{-1}\partial\epsilon({\bf k})/\partial k_\alpha$ and
$\sigma,\tau=\pm 1$.  In most situations the static optical conductivity
(${\bf q}=0, E=0)$ at zero temperature is of interest. Expression
\eqref{eq:real_conductivity} reduces in this case to
\begin{eqnarray}\label{eq:T0_conductivity}
 \mbox{Re}\ \sigma_{\alpha\beta}&=&\frac{e^2}{4\pi }\frac 1{N^2}\sum_{{\bf
      k},{\bf k}'} v_\alpha({\bf k})v_\beta({\bf k}')\sum_{\sigma\tau}
  (-\sigma\tau)  G({\bf k},E_F+i\sigma 0^+) G({\bf k},E_F+i\tau
  0^+) \left[\delta({\bf k}-{\bf k}')\right.\nonumber \\&& \hspace*{-20pt}\left.
    +\ \Gamma({\bf k},E_F+i\sigma 0^+,{\bf k},E_F+i\tau 0^+;{\bf k}'-{\bf k})
     G({\bf k}',E_F+i\sigma 0^+) G({\bf k}',E_F+i\tau 0^+)\right]\, .
\end{eqnarray}
We immediately see that the vertex function $\Gamma$ contributes to the
conductivity only if it depends on the transfer momentum
$\mathbf{k}'-\mathbf{k}$. It is due to the symmetry $\epsilon({\bf k})=
\epsilon(-{\bf k})$ and $v_\alpha({\bf k}) =-v_\alpha(-{\bf k})$.  Since
the CPA vertex does not depend on the transfer momentum, the vertex
corrections to the single-bubble electrical conductivity vanish in the CPA
treatment.

Eq.~(\ref{eq:T0_conductivity}) is not appropriate for approximate
calculations of the electrical conductivity.  The vertex corrections
contained in $\Gamma$ are {\em added} to the single-bubble conductivity so
that their negative contributions may reverse positive sign of the
conductivity, thus leading to unphysical behavior.

To avoid such a situation we represent the conductivity in a different
way.\cite{Janis00b} We utilize the Bethe-Salpeter equation in the
electron-hole channel to represent the vertex $\Gamma$ via the irreducible
one, $\Lambda^{eh}$.  Inserting its formal solution into
Eq.~(\ref{eq:T0_conductivity}) we obtain a new, equivalent representation
for the conductivity
\begin{eqnarray}
  \label{eq:cond_eh_repr}
  &&\mbox{Re}\ \sigma_{\alpha\beta}=
  \frac{e^2}{4\pi}\frac 1{N^2}\sum_{{\bf k},{\bf k}'} v_\alpha({\bf k})v_\beta({\bf
    k}')\sum_{\sigma\tau}  (-\sigma\tau)  G_\sigma({\bf k}) G_\tau({\bf k})
      \left\{1 -\left[\Lambda^{eh}_{\sigma\tau}G_\sigma
      G_\tau\right]\bullet\right\}^{-1} ({\bf
  k},{\bf k};{\bf k}'-{\bf k}) \ .
\end{eqnarray}
where we denoted $G_\sigma({\bf k})=G({\bf k},E_F+i\sigma 0^+)$ and
$\Lambda^{eh}_{\sigma\tau} = \Gamma({\bf k},E_F+i\sigma 0^+,{\bf
  k}',E_F+i\tau 0^+;{\bf q})$.  At least for not too strong disorder the
norm of the operator $\left\| \Lambda _{\sigma \tau }^{eh}G_{\sigma
    }G_{\tau }\right\| \lesssim 1$. Hence the conductivity remains in this
representation non-negative and free of spurious unphysical behavior.
However, unlike formula \eqref{eq:T0_conductivity}, representation
\eqref{eq:cond_eh_repr} is implicit and its application is conditioned by
our ability to solve the Bethe-Salpeter integral equation in the
electron-hole channel explicitly.

\subsection{Asymptotic expression in high dimensions}
\label{sec:cond_high_dim}

To be explicit in the assessment of the contributions of the two-particle
vertex to the electrical conductivity we again resort to the limit of high
spatial dimensions where we know the vertex and the self-energy explicitly.
We use a hypercubic lattice with a dispersion relation $\epsilon({\bf
  k})=-t/(2d)^{1/2} \sum_{\nu=1}^d \cos k_\nu$, where the group velocity is
$v_\alpha({\bf k})=t/(2d)^{1/2} \sin k_\alpha$. Further on we use an
analytic representation for the two-particle bubble in high dimensions
where it can be represented via a double Gaussian
integral\cite{Mueller-Hartmann89}
\begin{eqnarray}\label{eq:bubble_infty}
  \chi^{\pm}({\bf q};z_1,z_2)&=&
  -\mbox{sign}(\mbox{Im}z_1\mbox{Im}z_2)\!\int_{-\infty}^\infty
  \!\! d\lambda_1 d\lambda_2 \theta(\lambda_1\mbox{Im}z_1) \theta(\lambda_2
  \mbox{Im}z_2)\nonumber\\ &&\times
  \exp\left\{i\left(\lambda_1 x(z_1)   +\lambda_2 x(z_2)\right) \!-\!\frac
    14\left(\lambda_1^2+{\lambda}_2^2
      +2\lambda_1\lambda_2 X({\bf q})\right) \right\}
\end{eqnarray}
with $x(z)=z -\Sigma(z)$ and $X({\bf q})=\frac 1d\sum_{\nu=1}^d \cos(
q_\nu)$, and the Heaviside step function $\theta(x)$. Only parallel
conductivity survives on a hypercubic lattice and the integrals over the
Brillouin zone of squares of the velocity factorize, i.~e.,
\begin{eqnarray}
  \label{eq:velocity_hd}
  \sum_{\bf k} v_\alpha({\bf k})^2 G_\sigma({\bf k})  G_\tau({\bf k}+ {\bf
    q}) &=&   \frac {t^2}{2d} \chi_{\sigma\tau}({\bf q}) \, .
\end{eqnarray}

Using the above representations and simplifications in
Eqs.~\eqref{eq:2P_vertex_hd} and \eqref{eq:T0_conductivity} we obtain after
straightforward manipulations an explicit asymptotic formula for the
conductivity in high spatial dimensions
\begin{eqnarray}\label{eq:asymptotic_conductivity}
\frac{\pi }{e^2}\ \mbox{Re}\ \sigma_{\alpha\alpha}&=&\frac{t^2}{2d}\left\langle
  |\mbox{Im}G_+|^2\right\rangle -\frac{t^4}{16d^2}\ \mbox{Re}\left\{\frac{\langle
      G_+^2\rangle^4}{G_+^4}\left[\left \langle\frac 1{(1+G_+(\Sigma_+-V_i)
          )^2}\right\rangle_{av}-1\right]^2\right. \nonumber \\[4pt] && \left.
    \hspace*{-20mm} +\ \frac{\langle|G_+|^2\rangle^2}{|G_+|^2}
  \left[\left\langle\frac 1{|1+G_+(\Sigma_+-V_i)|^2}\right\rangle_{av}
    -1\right] \left( \frac{|\langle G_+^2\rangle|^2}{|G_+|^2}
  \left[\left\langle\frac 1{|1+G_+(\Sigma_+-V_i)|^2}\right\rangle_{av}
    -1\right] \right.\right.\nonumber \\[4pt]&&\left.\left.
-2\frac{\langle G_+^2\rangle^2}{G_+^2}  \left[\left\langle\frac 1{(1
      +G_+(\Sigma_+-V_i))^2}\right\rangle_{av} -1\right]\right)\right\}
\end{eqnarray}
where we used abbreviations $\langle G_+\rangle=N^{-1}\sum_{\bf k}G({\bf
  k},E_F+i0^+)$, $\langle G_+^2\rangle=N^{-1}\sum_{\bf k}G({\bf
  k},E_F+i0^+)^2$, and $\Sigma_+=\Sigma(E_F+i0^+)$. The one-particle
propagators are the CPA ones calculated with the self-energy from the local
approximation \eqref{eq:CPA_equation}. Conductivity
\eqref{eq:asymptotic_conductivity} depends explicitly only on the disorder
distribution and the dimensionality. The vertex contribution, proportional
to $t^4/d^2$, has negative sign and we cannot guarantee positivity of the
conductivity. Although the vertex corrections are negligible in the limit
$d\to\infty$ they may turn the overall sign of the conductivity negative
for a fixed finite dimension $d$. Whether conductivity
\eqref{eq:asymptotic_conductivity} for a fixed dimension is positive or
negative depends on the band filling and the disorder distribution and
strength. Vanishing of the asymptotic conductivity in this representation
does not indicate Anderson localization but merely a limit upon the
dimension below which formula \eqref{eq:asymptotic_conductivity} cannot be
applied.

\subsection{Mean-field expression for vertex corrections to the electrical
  conductivity }
\label{sec:mean_field_vcorr}

The limit of high spatial dimensions is usually used in order to set a
mean-field approximation for a physical quantity. We showed in the
preceding subsection that the vertex corrections are asymptotically less
important than the one-electron, single-bubble term. But the vertex
corrections may, in finite dimensions, turn static optical conductivity
negative.  Expression \eqref{eq:T0_conductivity} with the asymptotic
solution for the vertex function in high dimensions is hence unsuitable for
serving as a mean-field approximation, since physical consistence of the
result is not guaranteed.  We can take the leading single-bubble term as a
mean-field approximation for the conductivity as is actually common in the
literature.  Or, when we are interested in the impact of vertex
corrections, we have to use representation \eqref{eq:cond_eh_repr} and
evaluate its leading asymptotic behavior in high spatial dimensions as
suggested recently.\cite{Janis00b}

The limit of high spatial dimensions enables one to solve the
Bethe-Salpeter equation in the electron-hole channel explicitly. We
need only its leading asymptotic order.  Note that only the nonlocal
part of the vertex $\Lambda ^{eh}({\bf k},{\bf k};{\bf k}^{\prime
  }-{\bf k})$ having odd parity with respect to reflections in ${\bf
  k}$ and ${\bf k}^{\prime }$ contributes to the conductivity. We
hence have to take its leading asymptotics into account. We derive it
if we solve the Bethe-Salpeter equation in the electron-hole channel
for $\Lambda ^{eh}$ instead for $\Gamma$, i.e., $\Lambda^{eh}
=\Gamma\left\{ \bullet\left[ GG\Gamma\right] +1 \right\} ^{-1}$. Using
vertex $\Gamma$ from Eq.~(\ref{eq:d_infty_vertex}) one finds in the
order $O(1/d)$
\begin{eqnarray}
  \label{eq:decoupling}
\Lambda ^{eh}({\bf k}_1,z_1,{\bf k}_2,z_2;{\bf
  q})&=&\Lambda(z_1,z_2)\nonumber \\ &&\hspace*{-80pt}
+\left(1-\Lambda(z_1,z_2)G(z_1)G(z_2) \right)^2\left[\Gamma ({\bf
    k}_{1},z_{1},{\bf k}_{2},z_{2};{\bf q})- \Gamma^{eh} ({\bf k}_{2}- {\bf
    k}_{1};z_{1},z_{2})\right] .
\end{eqnarray}
In the limit $d\to\infty $ the momentum convolutions decouple. To derive
the leading asymptotic contribution from the nonlocal part of
$\Lambda^{eh}$ to the conductivity we have to calculate the momentum
convolutions at the level of order $O(1/d)$ so that the velocities appear
in squares and the momentum integrals do not vanish.  Keeping only the
leading-order terms in the expansion of the denominator in
Eq.~(\ref{eq:cond_eh_repr}) we end up with a mean-field-like expression for
the \emph{dc}-conductivity \cite{Janis00b}
\begin{subequations}\label{eq:mf_cond}
\begin{eqnarray}
  \label{eq:cond_positive}
  \mbox{Re}\ \sigma_{\alpha\alpha}=\frac{e^2}{4 \pi}
  \sum_{\sigma\tau}(-\sigma\tau) \frac{\langle v_\alpha^2 G_\sigma
    G_\tau\rangle }{1-\langle v_\alpha^2G_\sigma G_\tau\rangle
    \langle\Lambda^{\prime\alpha}_{\sigma\tau} \rangle }
\end{eqnarray}
where $\langle v_\alpha^2 G_\sigma G_\tau\rangle = N^{-1}\sum_{\bf
  k}v_\alpha({\bf k})^2 G_\sigma({\bf k})G_\tau({\bf k})$ and
\begin{eqnarray}
  \label{eq:Lambda_prime}
  \langle\Lambda^{\prime\alpha}_{\sigma\tau}\rangle&= &\frac
  1{N^{2}}\sum_{{\bf k},{\bf k}'}
  \frac{\delta^2} {\delta v_\alpha({\bf  k})\delta v_\alpha({\bf k}')}
  \Lambda^{eh}_{\sigma \tau}({\bf k},{\bf k};{\bf k}'-{\bf k}) \, .
\end{eqnarray}\end{subequations}
The irreducible vertex $\Lambda^{eh}$ is taken from
Eq.~\eqref{eq:decoupling}.  The one-particle propagators are those from the
local, coherent-potential approximation, since it is the local irreducible
vertex $\Lambda$ that determines the leading high-dimensional asymptotics
of the self-energy in Eq.~\eqref{eq:SE_vertex_im}.  However, only terms
with odd symmetry with respect to time inversion contribute to the vertex
corrections to the conductivity and we have to go to the leading order of
the nonlocal part of $\Lambda^{eh}$ to substantiate them.  If one resorts
to the local vertex $\Lambda$ in Eq.~\eqref{eq:mf_cond} one recovers the
CPA conductivity of the single electron-hole bubble.

Conductivity \eqref{eq:mf_cond} with the CPA one-particle propagator can be
called a mean-field approximation for the conductivity with vertex
corrections, since the vertex corrections are determined from the
asymptotic limit of high spatial dimensions. It can be applied in finite
dimensions for $d>2$ with the appropriate lattice structure determined by
the Brillouin zone. Note that the vertex $\Lambda^{eh}$ from
Eq.~\eqref{eq:decoupling} contains a diffusion pole from the
electron-electron channel and the integral in Eq.~\eqref{eq:Lambda_prime}
diverges in $d\le2$.  Hence the mean-field description of vertex
corrections break down there. In these low dimensions, where Anderson
localization is expected, we have to use the full two-particle
self-consistent parquet approximation in order to take properly into
account the influence of the diffusion pole.

\section{Binary alloy: high-dimensional approximation}
\label{sec:binary_alloy}

Mean-field expression for the conductivity with vertex corrections
\eqref{eq:mf_cond} demands the evaluation of momentum integrals over the
Brillouin zone to determine the averaged velocity and the derivative of the
vertex $\Lambda^{eh}$.  This must be done numerically for each particular
lattice. To assess qualitatively the influence of the vertex corrections on
the conductivity we can further simplify the mean-field expression in that
we approximate the momentum integrals by their high-dimensional
asymptotics, i.~e., we replace the momentum integrals by integrals with the
density of states. We use the same steps as in
Subsection~\ref{sec:vertex_high_dim} when deriving the high-dimensional
asymptotics of the conductivity. Using vertex \eqref{eq:decoupling} we
straightforwardly derive
\begin{eqnarray}
  \label{eq:cond_dos}
\mbox{Re}\ \sigma_{\alpha\alpha} &=& \left( \frac{e^{2}t^{2}}{8\pi
    d}\right) 
\sum_{\sigma\tau} \frac{(-\sigma\tau)  \langle G_\sigma G_\tau\rangle}
{1+\frac{t^{2}}{2d}   \langle G_\sigma G_\tau\rangle
  \Lambda_{\sigma\tau}\left(1- \Lambda_{\sigma\tau}G_\sigma G_\tau\right)
  \left[\gamma_{\sigma\tau}  \langle G_\sigma^2\rangle \langle G_\tau^2
    \rangle -\gamma_{\sigma\sigma} \langle G_\sigma^2\rangle^2
    -\gamma_{\tau\tau} \langle G_\tau^2\rangle^2\right]}\ ,
\end{eqnarray}
where $\langle G_{\sigma }G_{\tau }\rangle $ is defined as in
Eq.~\eqref{eq:asymptotic_conductivity}. The vertices $\Lambda$ and $\gamma$
are calculated in the local approximation from Subsection~\ref{sec:local}.
We can explicitly sum over the indices $\sigma,\tau$ and obtain
\begin{eqnarray}
  \label{eq:cond_dos_summed}
\mbox{Re}\ \sigma_{\alpha\alpha} &=& \left( \frac{e^{2}t^{2}}{4\pi
    d}\right) \left\{ \frac{ \langle |G_+|^2\rangle}
{1+\frac{t^{2}}{2d}   \langle |G_+|^2\rangle
  \Lambda_{+-}\left(1- \Lambda_{+-}|G_+|^2\right)
  \left[\gamma_{+-}  \left|\langle G_\sigma^2\rangle\right|^2
    -2\mbox{Re}\left(\gamma_{++} \langle G_+^2\rangle^2
    \right)\right]}\right.\nonumber  \\ &&\left. \hspace*{60pt} -
\mbox{Re}\ \frac{ \langle G_+^2\rangle}{1-\frac{t^{2}}{2d} \langle
  G_+^2\rangle^3 \Lambda_{++}^2}\right\} \ .
\end{eqnarray}

For the explicit numerical calculations we choose a binary alloy with the
site-diagonal disorder distribution
\begin{eqnarray}
  \label{eq:bin_alloy}
  \rho(V)&=& x\delta(V-\Delta) + (1-x)\delta(V+\Delta) \, .
\end{eqnarray}
To simplify the relation between the one and two-particle functions we
further choose a model density of states of a $d=\infty$ Bethe instead of a
hypercubic lattice. It is characterized by an equation
\begin{eqnarray}
  \label{eq:DOS_BL}
  G(z)&=& \frac 1{z-G(z)}
\end{eqnarray}
where we set the hopping $t=1$ with the energy band $E\in (-2,2)$. The
self-energy for complex energies in this model is then determined from a
cubic equation
\begin{subequations}\label{eq:1P_BA}
\begin{eqnarray}
  \label{eq:SE_BA}
  0&=& G(z)^3-2zG(z)^2+(1-\Delta^2+z^2)G(z)-(z-(1-2x)\Delta)
\end{eqnarray}
with
\begin{eqnarray}
\Sigma(z)&=& z-G(z)- G(z)^{-1} \, .
\end{eqnarray} \end{subequations}

We resort to real frequencies $E_\sigma=E+i\sigma0^+$ and take advantage of
explicit representations of two-particle functions via one-particle local
propagators
\begin{eqnarray}
  \label{eq:BA_G2}
  \langle G_\sigma G_\tau\rangle&=& \frac{G_\sigma G_\tau}{1-G_\sigma G_\tau}
\end{eqnarray}
and
\begin{eqnarray}
  \label{eq:BA_2PI_vertex}
  \Lambda_{\sigma\tau}&=& \frac 1{G_\sigma G_\tau} - \frac 1{\displaystyle
    \frac x{(E -G_\sigma-\Delta)(E -G_\tau-\Delta)} +\frac {1-x}{(E
      -G_\sigma+\Delta)(E -G_\tau+\Delta)}} \, .
\end{eqnarray}

We use the above formulas together with a numerical solution to the
self-energy from Eq.~\eqref{eq:1P_BA} in Eq.~\eqref{eq:cond_dos_summed} to
reach quantitative results for the electrical conductivity. In particular
we are interested in the impact of vertex corrections onto the CPA
conductivity.  We set the formal parameter of the lattice dimension $d=3$
in Eq.~\eqref{eq:cond_dos_summed}.

\begin{figure}[htb]
\epsfig{file=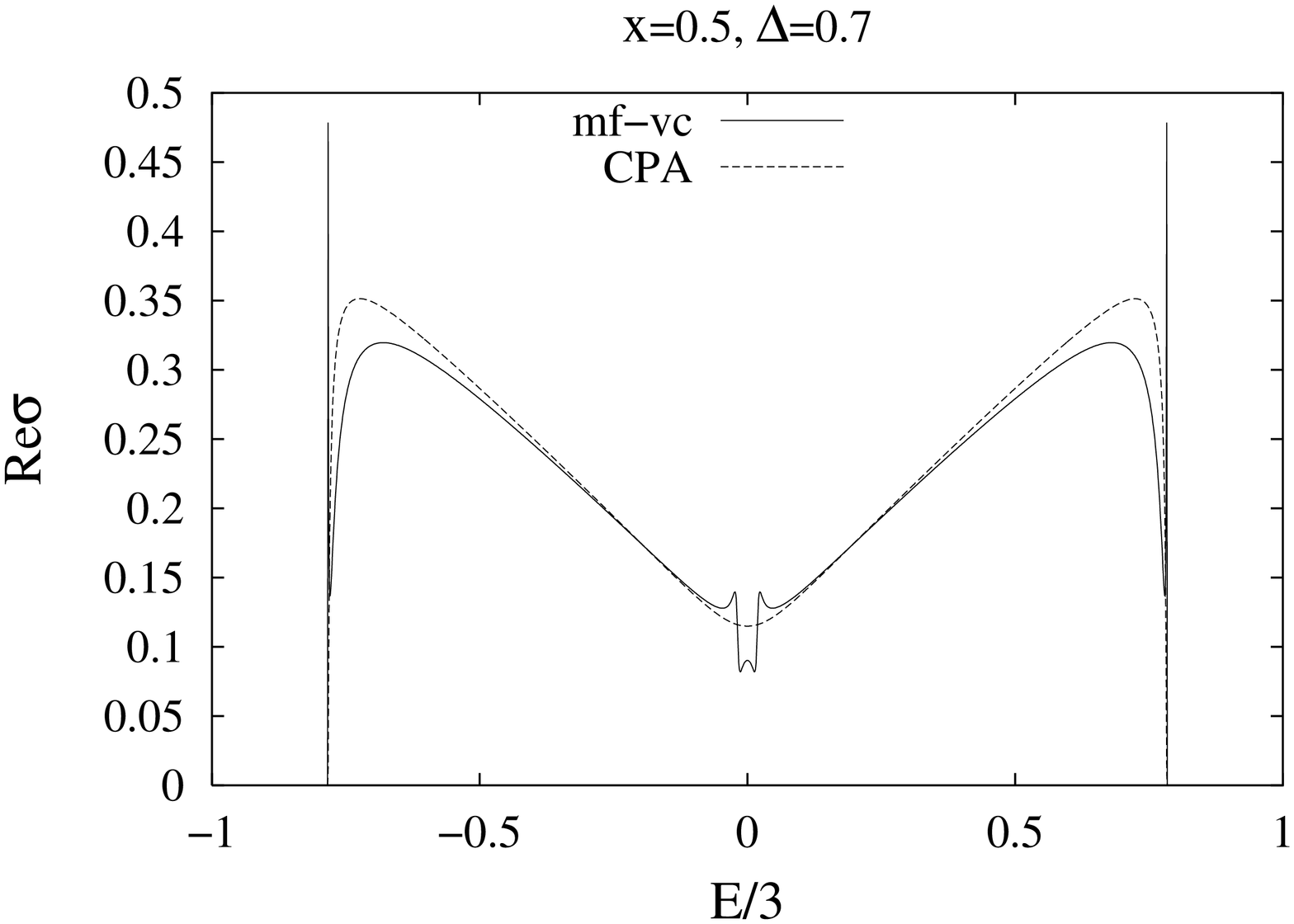, angle=-90, width=7.5cm} \epsfig{file=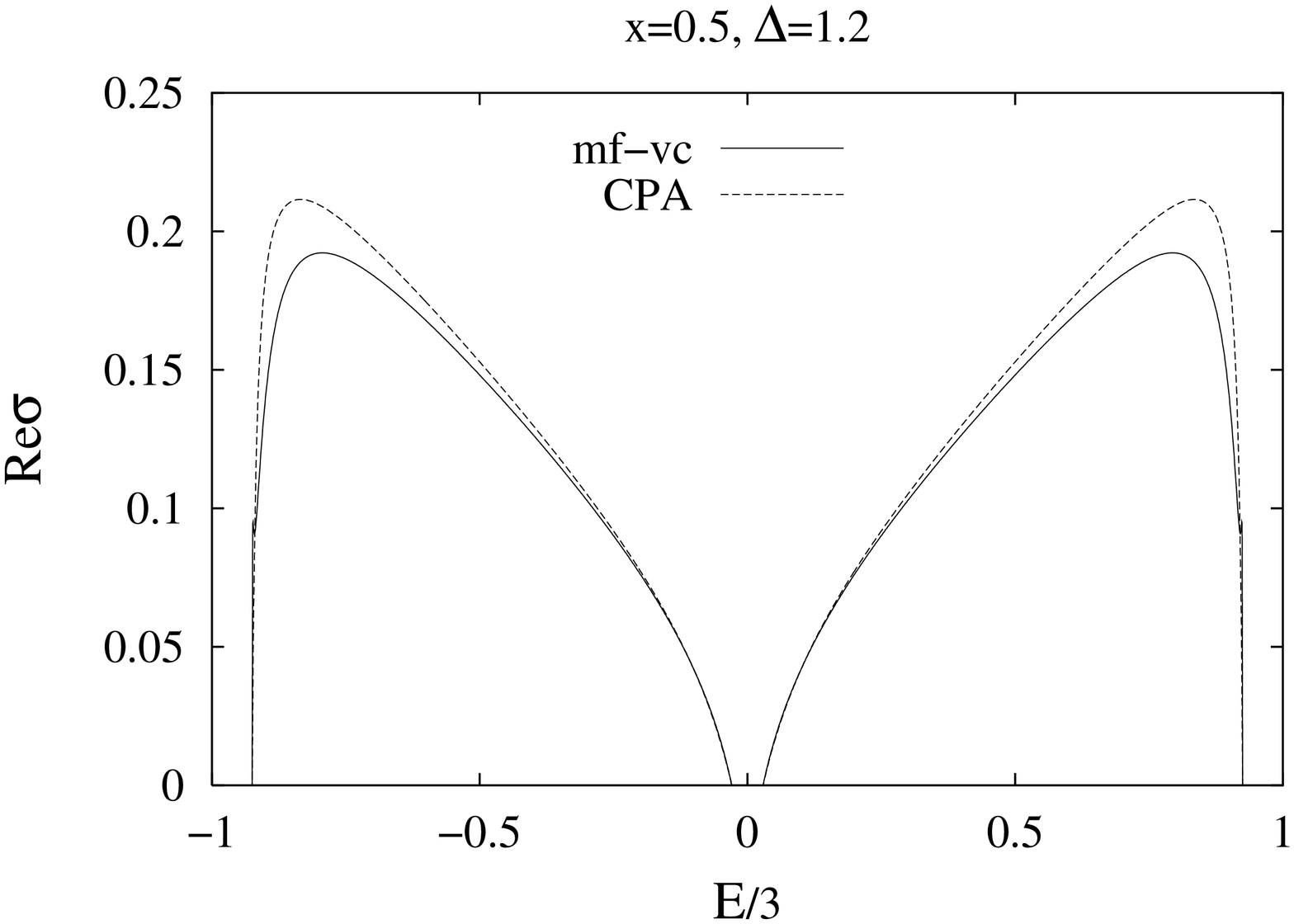,
  angle=-90, width=7.5cm}\medskip
\caption{Real part of the static conductivity as a function of energy for
  concentration $x=0.5$ and two values of disorder strength $\Delta=0.7$
  and $\Delta=1.2$.}
\label{fig:x0_5}
\end{figure}
\begin{figure}[htb]
\epsfig{file=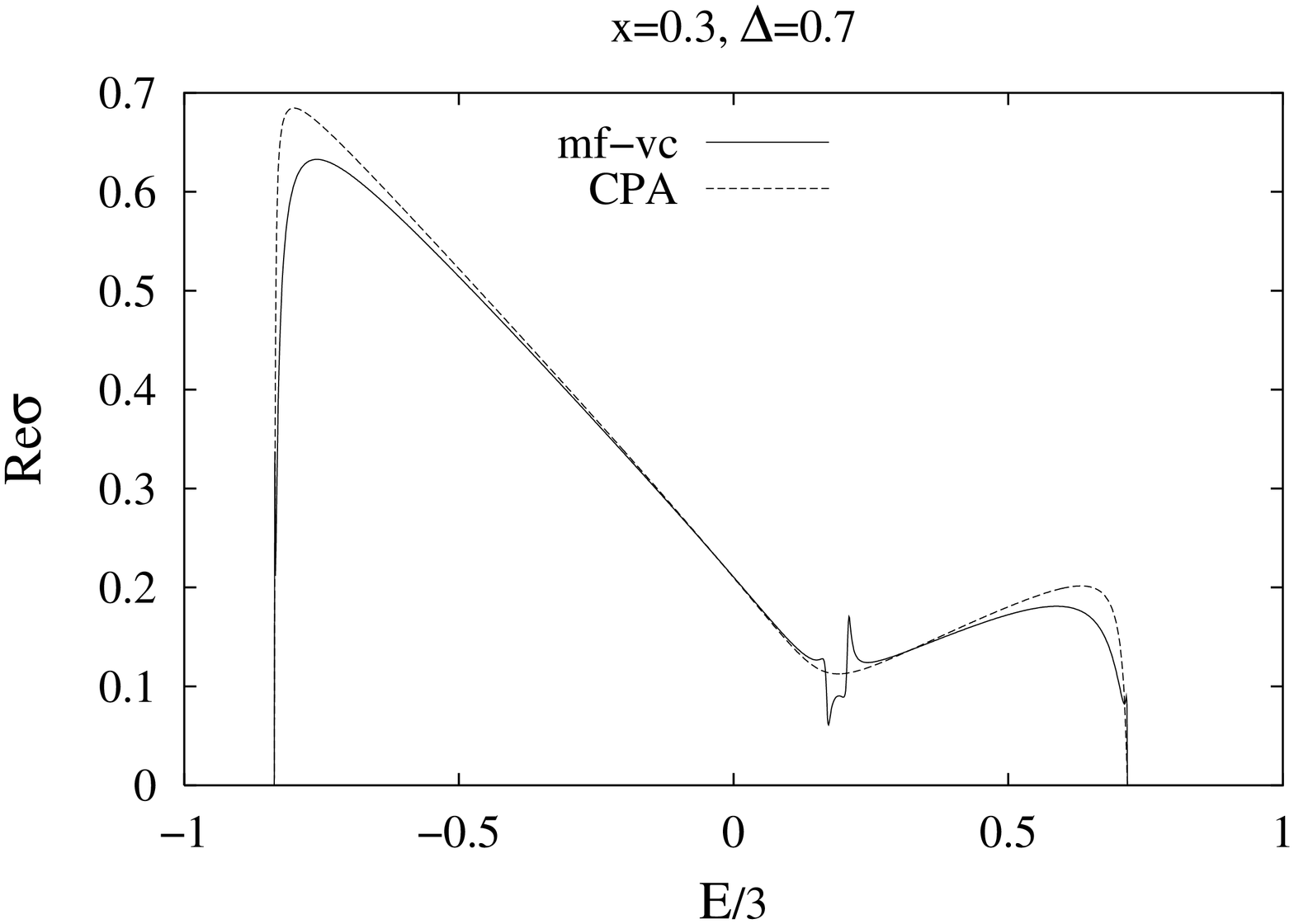, angle=-90, width=7.5cm} \epsfig{file=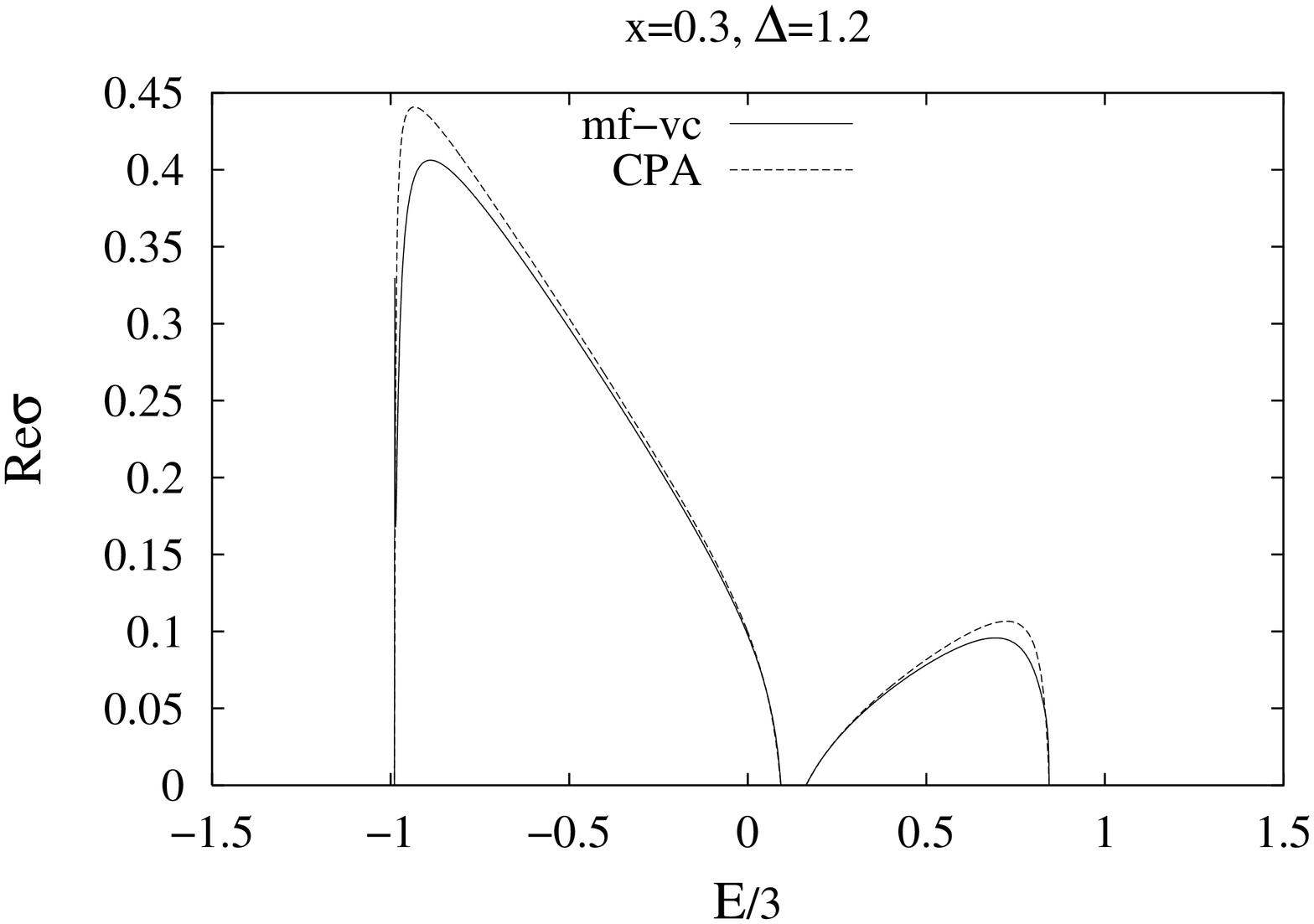,
  angle=-90, width=7.5cm}\medskip
\caption{Real part of the static conductivity as a function of energy for
  concentration $x=0.3$ and two values of disorder strength $\Delta=0.7$
  and $\Delta=1.2$.}
\label{fig:x0_3}
\end{figure}

Fig.~\ref{fig:x0_5} shows the CPA and the mean-field conductivity with
vertex corrections as a function of energy for a fixed concentration
$x=0.5$ and two values of the variance of the random potential (disorder
strength).  We see that the vertex corrections in general lower the
single-bubble conductivity. There are, however, situations where the vertex
corrections may lead to an increase in the conductivity. It is mostly due
to proximity of nonanalyticities in the local CPA vertex. The real part of
$\Lambda_{++}$ for $\Delta=1/\sqrt{2}$ goes through zero and displays a
pole for $x=0.5$.  This singular behavior causes sharp mobility peaks at
the band edges and irregularities near half-filling, $E=0$. Such a behavior
is observed only for $\Delta\approx 1/\sqrt{2}$ and $x\approx0.5$.
Fig.~\ref{fig:x0_3} shows the same for an asymmetric concentration $x=0.3$.
The mobility edges are now less pronounced and the conductivity
fluctuations inside the energy band are no longer symmetric.

\begin{figure}[htb]
\epsfig{file=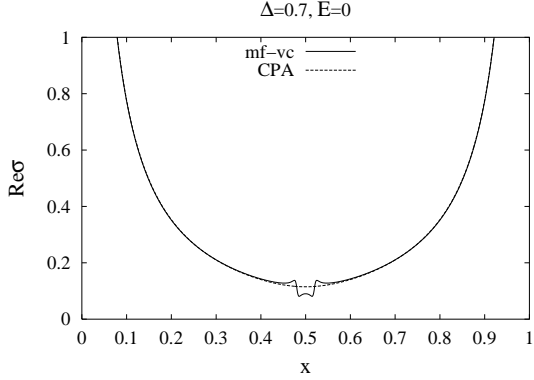, angle=-90, width=7.5cm}\medskip
\caption{Real part of the static conductivity as a function of
  concentration for the half-filled band and $\Delta=0.7$.}
\label{fig:D0_7}
\end{figure}

\begin{figure}[htb]
\noindent
\epsfig{file=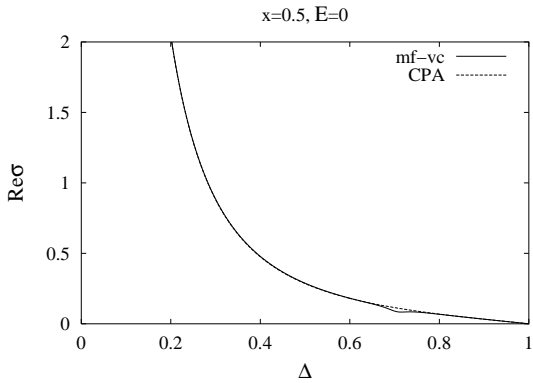, angle=-90, width=7.5cm}\medskip
\caption{Real part of the static conductivity as a function of disorder
  strength for the half-filled band and $x=0.5$.}
\label{fig:E0_0}
\end{figure}

Due to the structureless density of states of the Bethe lattice in
$d=\infty$, vertex corrections do not alter the CPA conductivity
significantly apart from the special situations influenced by the
singularity in the CPA vertex $\Lambda_{++}$.  Fig.~\ref{fig:D0_7} shows
the conductivity for the half-filled band as a function of concentration
$x$.  There is almost no significant difference between the conductivity
with and without vertex corrections, except for concentrations $x\approx
0.5$.  An analogous picture is obtained for the conductivity as a function
of the disorder strength $\Delta$, where the differences are significant
only around $\Delta\approx 1/\sqrt{2}$, Fig.~\ref{fig:E0_0}.

\section{Conclusions}
\label{sec:conclusions}

We have developed a diagrammatic theory for constructing systematic
approximations to nonlocal two-particle vertex functions of noninteracting
electrons moving in a random potential. The underlying idea of our approach
is to treat separately diagonal, local and off-diagonal, nonlocal elements
of the two-particle vertex. To this purpose we used the asymptotic limit of
high spatial dimensions. In the strict $d=\infty$ limit only the local
one-particle propagator is relevant and the solution contains only
single-site scatterings and reduces to the coherent-potential
approximation. Beyond this limit we utilized ambiguity in the definition of
the two-particle irreducibility. We classified the nonlocal contributions
to the two-particle vertex according to the type of the two-particle
irreducibility they belong to. Alike many-body theories there are three
topologically inequivalent irreducibility channels according to what pairs
of propagators interconnect spatially distinct two-particle scatterings on
the random potential. Representing the two-particle vertex via
Bethe-Salpeter equations and irreducible vertices in each channel and
utilizing the topological inequivalence of these representations we derived
a closed set of coupled (parquet) equations for the irreducible vertices.
The irreducible vertices from the parquet equations were used in an
integral form of the Ward identity to determine the self-energy of the
parquet solution. In this way we completed the parquet equations to an
approximation consistently determining all one and two-particle functions.
The input to the parquet approximation are the local CPA one-particle
propagator and two-particle vertex. Neither the form of the parquet
equations nor the Ward identity may cause unphysical nonanalyticities in
the solution. Solutions to the parquet approximation hence inherit the
analytic properties of the CPA input and are free of spurious, unphysical
behavior.

\begin{figure}[htb]
  \centerline{ \epsfig{file=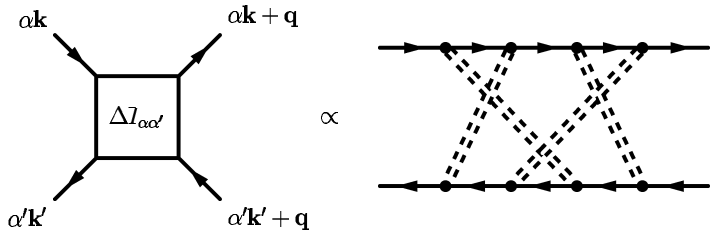, height=30mm}}\medskip
\caption{Lowest-order correction to the completely two-particle irreducible
  vertex $I_{\alpha\alpha'}=\gamma_{\alpha\alpha'}$ from the parquet
  approximation. The double-dashed line is the CPA local vertex
  $\gamma_{\alpha\alpha'}$. The internal fermion lines stand for the CPA
  off-diagonal propagator $G^{off}$.}
\label{fig:IR_corr}
\end{figure}

The proposed diagrammatic implementation of nonlocal corrections to the CPA
aims primarily at improving the CPA two-particle functions on a long-range
scale. Although there is no obvious small parameter controlling the
nonlocal corrections to the CPA, systematic improvements of the local
approximation are controlled via diagrams to the completely two-particle
irreducible vertex $I$. In the parquet approximation the input $I=\gamma$,
the local vertex from Eq.~\eqref{eq:local_vertex}. A first correction
$\Delta I$ to the input of the parquet approximation is proportional to
$\gamma^4(G^{off})^6$, where $G^{off}$ is the off-diagonal element of the
CPA propagator, see Fig.~\ref{fig:IR_corr}.  It means that in the
weak-disorder limit the parquet approximation is exact up to $V^7$ whereas
CPA only to $V^3$ in powers of the random potential. The parquet
approximation represents a significant systematic improvement of the CPA in
the weak-disorder limit.  We showed that beyond the weak-disorder limit the
parquet approximation contains the exact asymptotics of the two-particle
functions up to $d^{-2}$. This fact we used in proposing a mean-field
approximation for the electrical conductivity with vertex correction.

It is not, however, the weak-scattering limit where the assets of the
parquet approach to disordered systems lie. The parquet approach is in
particular appropriate for apprehending spatial quantum coherence and
backscattering effects. The parquet approximation even in the first
iteration (high-dimensional asymptotics) contains infinite number of
"crossed" two-particle diagrams. The weak localization with the diffusion
pole and a long-range spatial coherence from backscatterings are included.
Most importantly however, the parquet equations for the irreducible
two-particle vertex functions are fully self-consistent. They can
adequately deal with poles and divergences in the vertex functions and can
hence significantly change the diffusive character of the electronic
transport in low dimensions and strong disorder.

We hope that the presented parquet approach can bridge the gap between the
mean-field coherent-potential approximation on the one side and
localization theories on the other side.  Whether the parquet equations can
actually describe the localization transition must be decided by solving
the full set of self-consistent coupled integral equations for the vertex
functions and the self-energy.  This has been left for future research.

\section*{Acknowledgments}

It is my pleasure to acknowledge numerous stimulating and fruitful
conversations with D.~Vollhardt. I also benefited from discussions with
B.~Velick\'y, J.~Kudrnovsk\'y, V.~Drchal, and V. \v Spi\v cka on various
aspects of the electronic transport in random media.  The work was
supported in part by Grant No. 202/98/1290 of the Grant Agency of the Czech
Republic.

\begin{appendix}
\section{Algebraic derivation of Ward identities}

We use algebraic identities to prove equation \eqref{eq:W_BV} and the
special case of \eqref{eq:W_VW} with ${\bf k}_1={\bf k}_2$.

The two-particle function can be defined as a matrix element of a tensor
(direct) product $G^{(2)}_{ij,kl}(z_1,z_2) = \langle\langle ik|
\widehat{G}(z_1)\otimes\widehat{G}(z_2) | jl\rangle\rangle_{av}$ where the
resolvent operator is defined as $\widehat{G}(z)=\left[z\widehat{1} -
  \widehat{t}-\widehat{V}\right]^{-1}$ and the basis vectors are Wannier
orbitals at the lattice sites.  Using this lattice-space representation we
define a ``projection'' onto the one-particle subspace by equaling the
basis states from the left and right Hilbert space. When we sum over one
set of indices ($k=j$) we reduce the direct product from the two-particle
Hilbert space to an operator multiplication in the one-particle Hilbert
space. Ward identity (\ref{eq:W_BV}) is then a consequence of an operator
identity
\begin{eqnarray}
    \label{eq:operator_decomposition}
    \left[z_1\widehat{1} - \widehat{H}\right]^{-1}\cdot
      \left[z_2\widehat{1} - \widehat{H}\right]^{-1} &=& \left[z_2 -
      z_1\right]^{-1} \left\{\left[z_1\widehat{1} -
      \widehat{H}\right]^{-1} - \left[z_2\widehat{1} 
        - \widehat{H}\right]^{-1}\right\}
\end{eqnarray}
where $\widehat{H}=\widehat{t}+\widehat{V}$. Identity
\eqref{eq:operator_decomposition} holds for each configuration of the
random potential $\widehat{V}$ and after its configurational averaging we
obtain Eq.~(\ref{eq:W_BV}).  Ward identity (\ref{eq:W_BV}) is hence a
consequence of {\it completeness} of the eigenstates of the Hamilton
operator.  Completeness of the eigenstates reflects \textit{conservation of
  probability} and is a necessary prerequisite for conservation of energy
and other physical quantities.

To prove the Ward identity (\ref{eq:W_VW}) with ${\bf k}_1={\bf k}_2$ we
use the Bethe-Salpeter equation in the electron-hole channel
\begin{eqnarray}
G^{(2)}_{ij,kl}(z_1,z_2)&=& G_{ij}(z_1) G_{kl}(z_2) 
+ \sum_{i'j'k'l'} G_{ii'}(z_1) G_{ll'}(z_2)
\Lambda^{eh}_{i'j',k'l'}(z_1,z_2) G^{(2)}_{j'j,kk'}(z_1,z_2) .
\end{eqnarray}
We multiply it with the inverse one-particle propagators from left and
right and obtain
\begin{eqnarray}
  \label{eq:BS-multiplied}
\sum_{i'l'} G^{-1}_{ii'}(z_1)G^{-1}_{ll'}(z_2)
G^{(2)}_{i'j,kl'}(z_1,z_2)&=& \delta_{ij}\delta_{kl} 
 +\sum_{j'k'}\Lambda^{eh}_{ij',k'l}(z_1,z_2)
G^{(2)}_{j'j,kk'}(z_1,z_2) .
\end{eqnarray}
Summing over the intermediate indices and using Eq.~(\ref{eq:W_BV}) we find
\begin{eqnarray}
  \label{eq:W-multiplied}
\sum_{i'jl'}G^{-1}_{ii'}(z_1)G^{-1}_{ll'}(z_2) G^{(2)}_{i'j,jl'}
(z_1,z_2)&=& \frac 1{\Delta z}\left[G^{-1}_{il}(z_1)-
G^{-1}_{il}(z_2)\right]  .
\end{eqnarray}
We insert Eqs.~(\ref{eq:W-multiplied}) and (\ref{eq:W_BV}) in
Eq.~(\ref{eq:BS-multiplied}) and obtain the desired identity for the
self-energy
\begin{eqnarray}
\Delta\Sigma_{il}&=& \sum_{j'k'}\Lambda^{eh}_{ij',k'l}(z_1,z_2)
\Delta G_{j'k'} .
\end{eqnarray}
It reads in momentum space
\begin{eqnarray}
\Sigma({\bf k},z_1) -\Sigma({\bf k},z_2) &=& \frac 1{N}\sum_{{\bf k}'}
\Lambda^{eh}({\bf k},z_1,{\bf k},z_2;{\bf k}' -{\bf k})
\left[G({\bf k}',z_1) - G({\bf k}',z_2)\right] .
\end{eqnarray}
Analogously we can derive a Ward identity for the vertex function from the
electron-electron channel
\begin{eqnarray}
\Sigma({\bf k},z_1) -\Sigma(-{\bf k},z_2) &=& \frac 1{N}\sum_{{\bf k}'}
\Lambda^{ee}({\bf k},z_1,{\bf k}',z_2;{\bf k}' -{\bf k})
\left[G({\bf k}',z_1) - G(-{\bf k}',z_2)\right] .
\end{eqnarray}
\end{appendix}


\begin{thebibliography}{99}
\bibitem{Note2a} A two-particle function is called completely irreducible
  if it is irreducible in all two-particle channels, i.~e., it can be
  disconnected by cutting at least three lines (one-particle propagators).
\end{thebibliography}
\end{document}